\documentclass[showpacs,twocolumn,prx,superscriptaddress,floatfix]{revtex4-1}

\usepackage[utf8]{inputenc}
\usepackage{graphicx}
\usepackage{color}
\usepackage{amsmath}
\usepackage{amssymb}
\usepackage{subfigure}
\usepackage{epsfig}
\usepackage{float}

\usepackage{lipsum}

\newcommand{\sR}{\boldsymbol{\mathcal{R}}}
\newcommand{\J}{\boldsymbol{J}}
\newcommand{\F}{\boldsymbol{F}}
\newcommand{\D}{\boldsymbol{D}}
\newcommand{\B}{\boldsymbol{B}}
\newcommand{\pp}{\boldsymbol{p}}
\newcommand{\ppb}{\boldsymbol{\bar{p}}}
\newcommand{\pb}{\bar{p}}
\newcommand{\cc}{\boldsymbol{c}}
\newcommand{\bb}{\boldsymbol{b}}
\newcommand{\zz}{\boldsymbol{z}}

\newcommand{\vv}{\boldsymbol{v}}
\newcommand{\rr}{\boldsymbol{r}}
\newcommand{\xxi}{\boldsymbol{\xi}}

\newcommand{\eeta}{\boldsymbol{\eta}}
\newcommand{\ssig}{\boldsymbol{\sigma}}
\newcommand{\jj}{\boldsymbol{j}}
\newcommand{\ttau}{\boldsymbol{\tau}}
\newcommand{\ff}{f}

\definecolor{applegreen}{rgb}{0.0,0.5,0.0}

\begin{document}

\title{Lorentz forces induce inhomogeneity and fluxes in active systems}

\author{H.D.~Vuijk}
\affiliation{Leibniz-Institut  f\"ur Polymerforschung Dresden, Institut Theorie der Polymere, 01069 Dresden, Deutschland}
\author{J.U.~Sommer}
\affiliation{Leibniz-Institut  f\"ur Polymerforschung Dresden, Institut Theorie der Polymere, 01069 Dresden, Deutschland} \affiliation{Technische Universit\"at Dresden, Institut f\"ur Theoretische Physik, 01069 Dresden, Deutschland}
\author{H.~Merlitz}
\affiliation{Leibniz-Institut  f\"ur Polymerforschung Dresden, Institut Theorie der Polymere, 01069 Dresden, Deutschland}
\author{J.M.~Brader}
\affiliation{Department de Physique, Universit\'e de Fribourg, CH-1700 Fribourg, Suisse}

\author{A.~Sharma}
\email{sharma@ipfdd.de}
\affiliation{Leibniz-Institut  f\"ur Polymerforschung Dresden, Institut Theorie der Polymere, 01069 Dresden, Deutschland}\affiliation{Technische Universit\"at Dresden, Institut f\"ur Theoretische Physik, 01069 Dresden, Deutschland}


\begin{abstract}

We consider the nonequilibrium dynamics of a charged active Brownian particle in the presence of a space dependent magnetic field.
It has recently been shown that the Lorentz force induces a particle flux perpendicular to density gradients,
thus preventing a diffusive description of the dynamics.
Whereas a passive system will eventually relax to an equilibrium state,
unaffected by the magnetic field, an active system subject to a spatially varying Lorentz force settles into a
nonequilibrium steady state characterized by an inhomogeneous density and divergence-free bulk fluxes.
A macroscopic flux of charged active particles is induced by the gradient of the magnetic field only and does not require additional symmetric breaking such as density or potential gradients.
This stands in marked contrast to similar phenomena in condensed matter such as the classical Hall effect.
In a confined geometry we observe circulating fluxes, which can be reversed by inverting the direction of the magnetic field.
Our theoretical approach, based on coarse-graining of the Fokker-Planck equation, yields analytical results for the density, fluxes, and polarization in the steady state,
all of which are validated by direct computer simulation.
We demonstrate that passive tracer particles can be used to measure the essential effects of the Lorentz force on the active particle bath, and we discuss under which conditions the effects of the flux could be observed experimentally.
\end{abstract}


\maketitle

\section{Introduction}\label{sec:intro}

The Lorentz force modifies the trajectory of a charged particle without performing work on it.
In general the Lorentz force $\F_L$ acts on a moving charge $q$ as
\begin{equation}\label{F_Lorentz}
\F_L = q \vv \times \B~~, 
\end{equation}
where $\vv$ denotes the velocity vector of the charge and $\B$ is the applied magnetic field. The Lorentz force gives rise to important phenomena in solid-state physics such as the classical Hall effect,
in which electrical conductors develop a voltage difference transverse to the direction of the current if a magnetic field is applied perpendicular to the direction of the flow of charge carriers~\cite{hall1879on}.
Note that a macroscopic flux of charge carriers induced by Lorentz forces 
requires an explicitly broken symmetry that gives rise to a macroscopic velocity vector in addition to the symmetry breaking due to the magnetic field vector.
Simply speaking the classical Hall effect leads to a deviation of an existing macroscopic flux.
The main finding of this work is that in active systems driven by internal forces a macroscopic flux emerges from a flux-free system in the presence of a magnetic field gradient. 

While the effect of the Lorentz force on the properties of materials has been thoroughly studied in the context of solid-state physics \cite{altland2010condensed}, 
much less is known about its influence on soft-matter systems,
which are dominated by overdamped diffusive dynamics.
Part of the reason is that Lorentz forces may appear to be too weak to affect slowly moving colloidal particles significantly.
However, for Lorentz force to affect the motion of the particle, it is not its absolute value that is important but how its magnitude compares with the frictional force coming from the solvent.
Since both the forces depend on velocity, the strength of Lorentz force relative to frictional force can be quantified and shown to be significant for highly charged mesoscopic particles in strong magnetic fields. 


Before coming back to these points it is important to understand how the Lorentz force affects diffusive dynamics and how the diffusion equation can be extended to incorporate nonconservative velocity-dependent forces.
It is known that as a consequence of the Lorentz force the corresponding Smoluchowski equation for the probability distribution acquires a tensorial diffusion coefficient~\cite{balakrishnan2008elements},
the components of which are determined by the applied magnetic field and the friction coefficient.
Recently it has been shown that this tensor has an antisymmetric part,
which gives rise to additional fluxes in the direction perpendicular to density gradients~\cite{chun2018emergence, vuijk2019anomalous}.
In a passive system, these fluxes affect the dynamics of a system as it relaxes from an initial nonequilibrium state to an equilibrium state.
The equilibrium properties, however, are unaffected by the Lorentz force.
A known consequence of the Lorentz force is a reduction of the diffusion coefficient in the plane perpendicular to the magnetic field,
which can be understood by the deflection of the diffusing particle in the direction perpendicular to its trajectory according to the Eq. \eqref{F_Lorentz}.
This does not affect the equilibrium distribution of a system of passive particles.

A completely different situation arises when a driven system is subjected to an external magnetic field.
A particularly interesting example of a driven system is active matter, which is ubiquitous in biology; examples include molecular motors inside cells \cite{julicher1997modeling} and microscopic swimmers such as bacteria and algae \cite{cates2012diffusive}.
Besides biological active matter, there exists synthetic active matter, most notably Janus particles, which are 'two-faced' particles with only one of them functionalized. When the functionalized half of the particle is activated, chemically or under illumination, the particle is propelled forward~\cite{jiang2010janus,walther2013janus,buttinoni2012active,samin2015self,wurger2015self}.
In contrast to externally driven systems, active matter has the hallmark feature that it is out of equilibrium without an externally broken symmetry.
Much progress has been made in understanding the properties of active matter by using active Brownian particles (ABPs) as a model system,
which captures the essential features of an internally driven system in a minimalistic fashion.
ABPs violate time-reversal symmetry by consuming fuel to generate motion, often referred to as self-propulsion.
Besides the application to biological systems, active matter serves as a paradigm to study the effect of broken time-reversal symmetry and nonequilibrium steady states in general~\cite{katz1983phase,katz1984nonequilibrium,cates2012diffusive,fodor2016how}.

\begin{figure*}[t]
\vspace{-3cm}
\includegraphics[width=1.\textwidth]{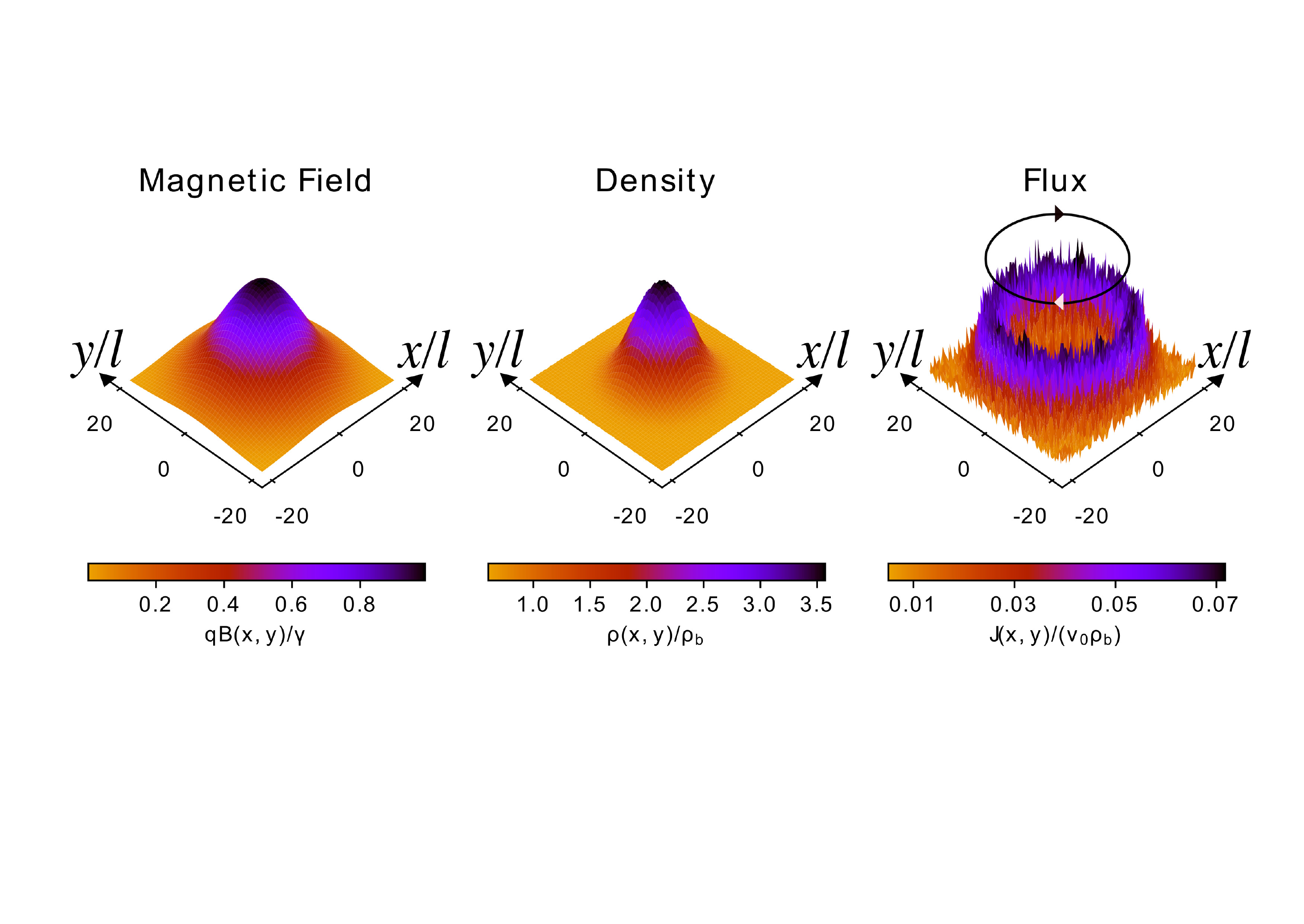}\vspace{-3.5cm}
\caption{Nonequilibrium steady state of an internally driven system (active Brownian particles) of noninteracting particles subjected to a Lorentz force arising from an external magnetic field.
$l$ denotes the persistence length of the ABP, $q$ its charge, $\gamma$ the friction coefficient, $v_0$ its self-propulsion speed, and $\rho_b$ is the bulk density.
The Gaussian shaped magnetic field is radially symmetric and points in the $z$ direction $B(x,y) \hat{\mathbf{e}}_z$.
The steady-state is characterized by (i) inhomogeneously distributed density and (ii) bulk fluxes.
The direction of the flux, indicated by the arrows, is perpendicular to the gradient of the magnetic field.
The flux can be reversed by  reversing the magnetic field.
In the absence of internal driving, there are neither fluxes nor density gradients in the steady state for any magnetic field.
This also holds for spatially homogeneous driving and no magnetic field.
However, when one considers the two together, homogeneously driven particles in an inhomogeneous magnetic field, the system exhibits the unusual nonequilibrium behavior.
The circulating fluxes are reminiscent of the Corbino effect in conductors~\cite{adams1915hall}.}\label{fig:gaussian}
The data was obtained from Brownian dynamics simulations of active Brownian particles subjected to Lorentz force (App. \ref{app:numerics}).
\end{figure*}

The absence of time-reversal symmetry can be exploited to obtain active fluxes by breaking the spatial symmetry of the system.
For instance, sliding motion of active particles along an asymmetric wall \cite{nikola2016active,katuri2018directed,rodenburg2018ratchet,reichhardt2017ratchet, dileonardo2009bacterial} can give rise to  fluxes along a boundary. 
 Similarly, in systems without walls or obstacles,
fluxes can be induced by making the activity spatially inhomogeneous \cite{stenhammar2016light}. 
 Explicitly including alignment interactions, such as torque on an active particle, can also give rise to fluxes in a spatially inhomogeneous activity field as has been shown experimentally in Ref. \cite{lozano2016phototaxis}. 
 Whereas in these studies activity has been time independent, one can induce fluxes by making the swim speed space and time dependent, as has been shown in theoretical and numerical studies \cite{geiseler2016chemotaxis,geiseler2017taxis,geiseler2017self,merlitz2018linear,zhu2018transport,lozano2019diffusing}.
In these systems the active particles 'surf' on an activity wave resulting in active fluxes. 

In this work, we show that the Lorentz force can induce macroscopic effects in a system of ABPs such as a magnetic-field dependent density distribution and flux-patterns.
Note that, in contrast to externally driven system (for example, charge carriers in conductors under an applied voltage), ABPs do not have a preferential direction of motion.
Thus, inhomogeneities are solely controlled by the spatial variation of the magnetic field.
Whereas interaction between particles were necessary to induce bulk fluxes in a system with a space-dependent swim speed which is constant in time \cite{stenhammar2016light}, this is not the case for the system studied here.
Moreover, with Lorentz forces, the fluxes can be reversed by reversing the direction of the magnetic field.

For slowly moving particles, Lorentz forces can be extremely small.
However, the effect of the Lorentz force on the motion of a particle is determined by ratio of the magnitudes of the Lorentz force and the frictional force. This ratio, which we refer to as the diffusive Hall constant, may become significant for highly charged particles in strong magnetic fields. 
We note that ABPs with magnetic interaction have been studied before; however, only by including a magnetic dipole moment without considering the Lorentz force.
\cite{vidal-urquiza2017dynamics,weber2011active,
radtke2012directed,ao2014active,
kline2005catalytic}.

As an example of the effect of the Lorentz force on charged active particles we show the results of Brownian dynamics simulations where the ABPs are subjected to an inhomogeneous magnetic field in Fig. \ref{fig:gaussian}.
The Gaussian shaped magnetic field is radially symmetric.
The two signatures of the Lorentz force on ABPs are clearly visible: Circulating fluxes perpendicular to
the gradient of the magnetic field and an inhomogeneous density distribution of the particles.
The fluxes observed in ABPs under Lorentz force are reminiscent of the Corbino effect in electrical conductors \cite{adams1915hall}.
Whereas a radial flux of charges is necessary for the Corbino effect in conductors, this is not the case for ABPs
which do not require explicit symmetry breaking in the system except the spatial inhomogeneity of the magnetic field.

On a microscopic level, fluxes in our system are related to the polarization of the active particles.
Polar ordering can emerge in active systems without aligning interactions. 
This has been shown in previous studies, for instance, under gravity \cite{enculescu2011active,stark2016swimming} and inhomogeneous activity \cite{stenhammar2016light,sharma2017brownian}. 
As in these studies, the origin of polarization in our system is kinetic in nature.
However, in contrast to previous studies, the particles in our system orient not only along the gradient of the applied field but also perpendicular to it.
Related to this polar order are active fluxes in the same direction.

A common starting point of the theoretical description of ABPs is the Langevin equation. 
Most models of ABP ignore inertial effects and are based on overdamped equations of motion~\cite{sharma2016green-kubo,sharma2017brownian,cates2013when,marconi2015towards,vuijk2018pseudochemotaxis,merlitz2018linear}.
Due to the nonconservative nature of the Lorentz force, the overdamped equation of motion cannot be obtained by simply setting the mass of the particles to zero~\cite{lau2007state, chun2018emergence}.
Though there exists a nontrivial limiting procedure~\cite{hottovy2012noise,hottovy2015smoluchowski,volpe2016effective} that yields the small-mass limit of the (velocity) Langevin equation, the resulting overdamped equation is not suitable for determining velocity dependent variables such as flux and entropy~\cite{celani2012anomalous, marino2016entropy, birrell2018entropy, chun2018emergence, vuijk2019anomalous} despite the fact that it captures the position statistics accurately.
In this work, we have avoided these problems by performing simulations using the (velocity) Langevin equation with a finite but small mass.
Our theoretical approach is based on coarse-graining of the Fokker-Planck equation using a gradient expansion~\cite{schnitzer1993theory,cates2013when}
, which yields an equation for the flux and density distribution as a function of time and the position variable alone.
This equation also yields the space-dependent polarization in the system by identifying the corresponding contribution from the orientational degrees of freedom.
We further show that one can improve the accuracy of the results by combining the coarse-graining approach with linear-response theory.
The results of this method match those of Brownian dynamics simulations. 

Our calculations lead to a straightforward estimation of the magnitude of this new kind of Hall effect in ABPs by comparing the amount of directed motion with the undirected motion of the active particles.
We present a possible experimental realization of our findings: The motion of a large tracer particle, insensitive to Lorentz force, would become anisotropic due to the presence of the (active) flux patterns. Using reasonable estimates for the system parameters, our calculations suggest that this anisotropy could be measured experimentally.


\section{Active particles with Lorentz forces}\label{sec:active_lorentz}

We consider a single spherical, self-propelled, charged Brownian particle in a magnetic field $\B(\rr)$ with a spatially varying magnitude but constant direction.
The active motion of the particle is modeled as a force in the direction of the orientation of the particle specified by an embedded unit vector $\pp$ undergoing rotational diffusion.
The state of the particle is determined by the position coordinate $\rr(t)$, the velocity $\vv(t)$ and 
the orientation $\pp(t)$. 

The dynamics are described by the following (Stratonovich) stochastic differential equations:
\begin{equation}\label{sde_r}
\frac{d \rr(t)}{dt} = \vv(t),
\end{equation}
\begin{align}\label{sde_v}
m\frac{d \vv(t)}{dt} = -\gamma \Gamma(\rr(t)) \cdot \vv(t) + \ff \pp(t)+ \sqrt{2\gamma T} \xxi(t),
\end{align}
and
\begin{equation}\label{sde_p}
 \frac{d\pp(t)}{dt} = \sqrt{2 D_r} \eeta(t) \times \pp(t),
\end{equation}
where  $m$ is the particle mass, $\gamma$ is the friction coefficient, $\ff$ is the self-propulsion force, 
$T$ is the temperature in units such that Boltzmann constant is equal to one and
$D_r$ is the rotational-diffusion constant.
The (dimensionless) matrix $\Gamma(\rr)= \boldsymbol{1} + \frac{qB(\rr)}{\gamma}M$, where $q$ is the charge of the particle and 
$M$ is a matrix such that $B(\rr) M \cdot \vv = \B(\rr) \times \vv$, and $B(\rr) \equiv |\B(\rr)|$.
The stochastic vectors $\xxi$ and $\eeta$ are Gaussian distributed with zero mean and autocorrelation
$\left<\xxi(t)\xxi^T(t')\right> =  \left<\eeta(t)\eeta^T(t')\right> =  \delta(t-t') \boldsymbol{1} $.
Note that there is no direct coupling between the orientation and the magnetic field.

We start with Eq.~\eqref{sde_v}  with the inertia term
because in presence of a Lorentz force,
the overdamped equation of motion cannot be used to determine observables depending on the velocity process (such as flux),
even though the position-statistics can be accurately described~\cite{hottovy2015smoluchowski,chun2018emergence,vuijk2019anomalous}.
However, the Fokker-Planck equation (FPE) for the overdamped motion corresponding to Eqs. \eqref{sde_r} and \eqref{sde_v} can be obtained by a different method based on an expansion in powers of mass.
This expansion removes the velocity degrees of freedom  (see App. \ref{app:small_mass_limit}) and yields the following FPE for the probability density $Q(t) \equiv Q(\rr,\pp,t)$:
\begin{align}\label{FPE_Q}
 \frac{\partial}{\partial t} Q(t) &=  -\nabla_{\rr} \cdot \J(\rr,\pp,t)+  D_r \sR \cdot \J_{rot}(\rr,\pp,t)
\end{align}
 where 
\begin{align}
 \J(\rr,\pp,t) = -  \frac{1}{\gamma} \Gamma^{-1} \cdot \left( T \nabla_{\rr} -  
  f \pp \right)Q(t),
\end{align} 
is the translational flux,
\begin{align}
\J_{rot}(\rr,\pp,t) = -D_r \sR Q(t)
\end{align}
is the rotational flux,
$\sR=\pp\times\nabla_{\pp}$ is the rotation operator, and

\begin{align}\label{Gamma_inv}
\Gamma^{-1}(\rr) = \boldsymbol{1} - \frac{\kappa(\rr)}{1+\kappa^2(\rr)} M
+   \frac{\kappa^2(\rr)}{1+\kappa^2(\rr)} M^2,
\end{align}
with 
\begin{align}\label{kappa}
\kappa(\rr) = \frac{q B(\rr)}{\gamma}~~.
\end{align}
This parameter, which we refer to as the diffusive Hall constant, measures the strength of the Lorentz force relative to the frictional force in the system.
When this parameter is comparable to unity, the Lorentz force affects the motion of the particle significantly.
As we show later in Sec.~\ref{sec:passive_diff}, the Lorentz force may become comparable to the frictional force for highly charged particles in strong magnetic fields.

Note that the matrix $\Gamma^{-1}$ in Eq.~\eqref{Gamma_inv} is not symmetric, and therefore, $T\Gamma^{-1}/\gamma$  cannot be interpreted as a diffusion tensor~\cite{vuijk2019anomalous}.
Due to the antisymmetric terms in $\Gamma^{-1}$, there are additional fluxes in the system that are perpendicular to the density gradient.

Equation~\eqref{FPE_Q} can be coarse grained further by a gradient expansion to remove the orientational degrees of freedom.
We consider the ansatz
\begin{align}\label{grad_exp_ansatz}
Q(\rr,\pp,t) = \rho + \ssig \cdot \pp + 
\ttau: \left(\pp \pp - \boldsymbol{1}/3\right) +
\Omega,
\end{align}
where the scalar $\rho$, the vector $\ssig$ and the matrix $\ttau$ are function of $\rr$ and $t$, and $\Omega$ contains all higher-order terms.
Equations for the time evolution of the probability density $\rho$, the total orientation $\ssig/3$ and higher-order tensors can be obtained by inserting Eq.~\eqref{grad_exp_ansatz} into Eq.~\eqref{FPE_Q} and integrating out the orientation degrees of freedom (see App. \ref{app_gradient_expansion}).
In the limit where gradients are small compared to the persistence length of the ABP~\cite{schnitzer1993theory,cates2013when}, one obtains a FPE for the probability density $\rho(\rr,t)$ as a function of time and position alone:
\begin{align}\label{FPE_rho}
 \frac{\partial}{\partial t} \rho(\rr,t) =
 - \nabla \cdot \jj(\rr,t),
\end{align}
where
\begin{align}\label{j}
\jj(\rr,t) =   \Gamma^{-1}(\rr) \cdot
	\left[ \frac{\ff}{\gamma} \ppb(\rr,t) \rho(\rr,t) \right] 
	- \frac{T}{\gamma} \Gamma^{-1}(\rr) \cdot \nabla \rho(\rr,t),
\end{align}
and

\begin{align}\label{polarization}
\ppb(\rr,t) = - \frac{l}{3\rho(\rr,t)}  \nabla \cdot \left[ \Gamma^{-1}(\rr) \rho(\rr,t) \right],
\end{align}
where 
\begin{align}\label{persistence}
l=\frac{\ff}{2 D_r \gamma}
\end{align}
is the persistence length of the ABP,
 and $\ppb(\rr,t)$ is the polarization, which is defined as the average orientation per particle
$ \left< \pp \delta \left( \rr' - \rr \right) \right>/ \rho(\rr,t)  $
where the angle brackets indicate an average with respect to the probability density of $\rr'$, $\pp$ and $\vv$ at time $t$.

\section{Nonequilibrium Steady State}\label{sec:noneq_ss}

Here we focus on the steady-state bulk behavior of the system and therefore use periodic boundary conditions in all directions.
Furthermore, the analysis is restricted to a magnetic field with a constant orientation, in this case the $z$ direction, and a magnitude that depends on a single coordinate, say $x$. 
Since the motion along the direction of the magnetic field is not affected by the Lorentz force, there is neither flux nor density variation along the $z$ direction in steady state.
For the chosen magnetic field, the system is translationally invariant in the $y$ direction; therefore, the flux and the polarization in the $y$ direction are independent of the $y$ coordinate.
This also implies that in steady state there can be no flux in the $x$ direction  
because of the periodicity and equivalence of the positive and negative $x$ directions.
Moreover, there can be no local $x$ fluxes forming closed loops via the $y$ direction because of the translation invariance in the $y$ direction.
The absence of fluxes in the $x$ direction has been corroborated by simulations.

By setting the $x$ component of the flux [Eq.~\eqref{j}] to zero, we obtain the following analytical expression for density:
\begin{align}\label{rho}
\rho(x) = N \left[ 1 +  \kappa^2(x) \right]^{\delta/2},
\end{align}
where $N^{-1}=\int_0^L dx \left[ 1 +  \kappa^2(x) \right]^{\delta/2}$
and 
\begin{align}\label{delta}
\delta =\frac{\Delta T}{T+ \Delta T}~~,
\end{align}
where $\Delta T = \ff^2/6D_r \gamma$.
%
%
The polarization can be calculated by inserting Eq. \eqref{rho} in Eq.~\eqref{polarization}, which yields
\begin{align}\label{px}
\pb_x(x) = \frac{ ( 2 - \delta)l}{3} \frac{ \kappa(x) \kappa'(x)}
{ \left[ 1 + \kappa^2(x) \right]^2},
\end{align}
for the $x$ component, and
\begin{align}\label{py}
\pb_y(x) = - \frac{l}{3} \frac{\kappa'(x)}{ \left[
1 + \kappa^2(x) \right]^2} \left[ 1 -
(1-\delta) \kappa^2(x)  \right]
\end{align}
for the $y$ component,
where $\kappa'(x) = \frac{d \kappa(x)}{d x}$.

The flux has two contributions, one coming from the polarization and the other from the gradient in the density [see Eq. \eqref{j}].
In steady state, the two contributions to the flux in the $x$ direction are equal in magnitude, but in opposite direction, resulting in zero flux in this direction.
In the $y$ direction there can be no density gradients because of the translation symmetry in this direction.
This means that the contribution to the flux in the $y$ direction coming from a density gradient along this direction is zero; however, the contribution from the polarization is not zero, resulting in a flux in the $y$ direction.
This flux can be calculated by using the solution for density [Eq. \eqref{rho}] as input to the $y$ component of Eq. \eqref{j}, yielding 
\begin{align}\label{yflux}
j_y(x) = \rho(x) \frac{\ff}{\gamma} \pb_y(x).
\end{align}
Note that the orientation and flux in the $y$ direction are perpendicular to the magnetic field gradient.

Before we test the predictions of the theory, a few remarks are in order.
First, $\Delta T$ appearing in Eqs.~\eqref{rho}--\eqref{py} has units of 
energy (note that the Boltzmann constant is unity, so $T$ has units of energy as well) and has previously been interpreted as an effective increase in the temperature due to activity~\cite{solon2015pressure, farage2015effective}.
However, we do not interpret $\Delta T$ as increase in temperature because this system cannot be mapped to an equilibrium system at a higher temperature.
Second, the expressions for the density, flux and polarization can be improved significantly by combining the gradient expansion with linear-response calculations of the polarization.
Such a calculation results in a differential equation for the polarization, which can be solved numerically.
This solution can be used to replace the linear-order contribution of the activity to the polarization calculated using the gradient expansion and using this polarization as input to the flux in Eq. \eqref{rho}.
Details of the linear-response theory, its application to the polarization and the combined solution are shown, respectively, in Appendices \ref{app:linear_response}, \ref{app:orientation} and \ref{app:improved_solution}.
Finally, it is interesting to note that $j_y(x)/\rho(x)$ is the macroscopic velocity of the ABP in the presence of the gradient field, while $\ff/\gamma$ gives the intrinsic microscopic velocity of the ABP.
While the intrinsic velocity leads to random motion on large time scales, the field-induced motion persists (in our case in the $y$ direction) on any time scale.
Thus the polarization  [Eq. \eqref{py}] gives the fraction of the swim force that is converted into macroscopic directed motion.
\begin{figure*}
\centering
\includegraphics[width=2.\columnwidth]{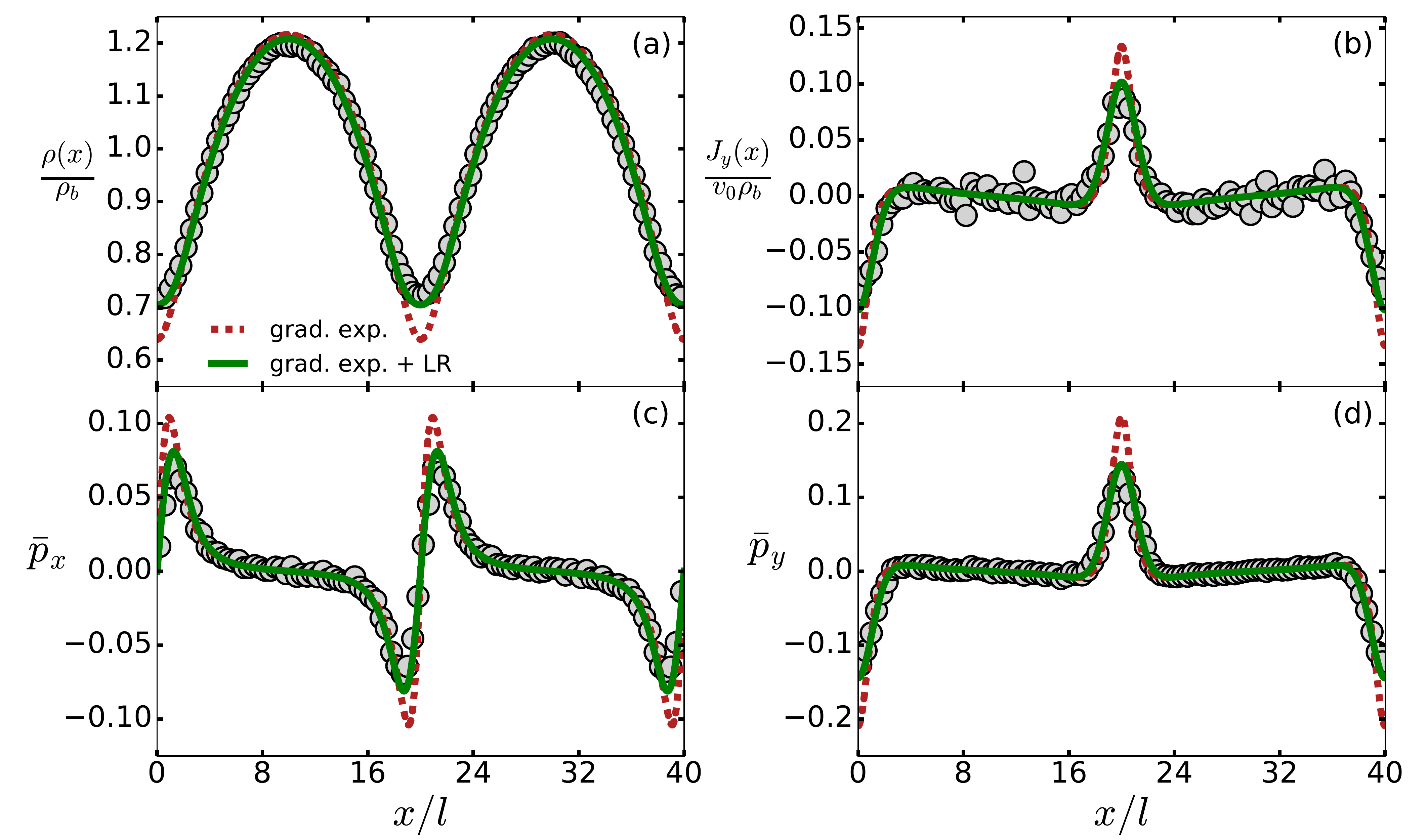}
\caption{(a) Density, (b) flux, and (c) and (d) orientation for $\ff=10$, $\kappa(x)=4 \sin(2\pi x /10)$, $D_r=20$, $\gamma = 1$ and $m=0.002$.
The self-propulsion speed of the ABPs is $v_0=\ff/\gamma$, $l$ is their persistence length, and $\rho_b$ is the bulk density.
The red, dashed lines correspond to Eqs.  \eqref{rho}--\eqref{yflux}.
The green, solid lines are the improved solutions, which replaces the lowest order term in $\ff$ in the solutions with the linear-response results  (see App. \ref{app:improved_solution}).
The symbols represent the results from Brownian dynamics simulations (see App. \ref{app:numerics}).
(a) The density is high where the magnitude of the magnetic field is large.
This corresponds to accumulation in regions where the diffusion is hindered.
(b) The flux in the system is proportional to the density and the polarization in the $y$ direction [see Eq. \eqref{yflux}].
There is flux in negative $y$ direction around $x=0$ and $x=40l$,
and in the positive direction around $x=20l$.
Note that, due to the periodicity of the system, the integral of the flux over the $x$ coordinate is zero and there is no center-of-mass motion.
From (c) and (d) it is clear that the polarization is nonzero in a narrow space interval and changes rapidly in this interval;
for example, from $x\approx 18l$ to $x \approx 22l$ the polarization rotates clockwise from the negative to the positive $x$ direction.
The analytical predictions from the gradient expansion overestimate the change in density, the flux and the orientation;
the prediction obtained from the combination of the gradient expansion and linear response theory, however, has good agreement with the simulation results.
}\label{fig:bulk_results}
\end{figure*}

Figure \ref{fig:bulk_results} shows the predictions of the theory for the density, flux and polarization for $\kappa(x) = 4\sin(2\pi x/10)$ and $\gamma = 1$ together with the results from Brownian dynamics (BD) simulations. 
Consistent with the prediction of equation \eqref{rho}, the density is high where the magnitude of the magnetic field is large (see Fig. \ref{fig:bulk_results} a). Qualitatively, the accumulation of particles in the high magnetic field region can be understood as follows. 
A magnetic field hinders the diffusive motion of Brownian particles because there is a force perpendicular to the velocity which results in circular motion.
For passive Brownian particles this means that on time scales longer than the velocity autocorrelation time the effect of a uniform magnetic field is a decrease of the diffusion constant in the plane perpendicular to the magnetic field \cite{balakrishnan2008elements}.
For a spatially varying magnetic field, the diffusion coefficient of particles is small in the region where magnetic field is large.
Whereas, for passive particles, the steady-state density distribution is independent of the diffusion coefficient, active particles accumulate in the region of low swim speed, which correspond to regions of high friction, or in this case high magnetic field~\cite{cates2012diffusive,sharma2017brownian}.

Passive Brownian particles in a space dependent magnetic field have a uniform steady-state density (a Boltzmann distribution with a uniform potential).
A system of ABPs with a homogeneous activity also has a uniform steady-state density.
However, the combination of a space-dependent magnetic field and uniform activity results in an inhomogeneous steady-state density.
An inhomogeneous density without a space dependent potential means that the density does not follow a Boltzmann distribution, a hallmark of nonequilibrium systems.

There is polarization in the plane perpendicular to the magnetic field;
see Eqs. \eqref{px} \eqref{py} and Figs. \ref{fig:bulk_results} c and d.
In steady state, the polarization and the gradient of the density in the $x$ direction are such that the fluxes resulting from them cancel each other.
The polarization in the $y$ direction, however, cannot be canceled by a density gradient because when the magnetic field varies only along the $x$ direction, the system is translation invariant in the $y$ direction and therefore, the density, flux and polarization cannot depend on the $y$ coordinate. 
Because there is polarization in the $y$ direction, but no density gradient to counter the flux from the polarization, there is in steady state a flux in the $y$ direction [see Eq. \eqref{yflux} and Fig. \ref{fig:bulk_results} b].
Due to the periodicity of the field, there is flux in both the positive as the negative  $y$ direction resulting in zero center-of-mass movement. 

Note that there is no direct coupling between the magnetic field and the polarization [see Eqs. \eqref{sde_v} and \eqref{sde_p}], and the inhomogeneous density along the $x$ direction is not a \emph{result} of the polarization.
Both the polarization and the inhomogeneous density are kinetic in nature, and result from a filtering mechanism.
The magnetic field hinders the motion of the ABP and a particle with an orientation parallel to the gradient of the magnetic field will get stuck.
A particle with an orientation antiparallel to the gradient of the magnetic field can more easily move towards regions of lower magnetic field.
These effects result in a force imbalance between regions of small and large magnetic fields.
In steady state the force imbalance is canceled by gradient in the chemical potential due to the increase of density in the high magnetic region. 

The presence of steady-state bulk fluxes is a key feature of the nonequilibrium steady state induced by Lorentz forces in a system of active particles.
These fluxes are unique in the sense that the underlying mechanism is independent of inter-particle interactions~\cite{stenhammar2016light}, aligning torques~\cite{lozano2016phototaxis} or sliding motion along asymmetric walls~\cite{nikola2016active,katuri2018directed,rodenburg2018ratchet,reichhardt2017ratchet, dileonardo2009bacterial}.
Moreover, the direction of fluxes is perpendicular to the gradient of the applied magnetic field.
This particular property is what gives rise to the circulating steady-state fluxes shown in Fig.~\ref{fig:gaussian} in which the magnetic field is radially symmetric.
The direction of the fluxes can be reversed by simply reversing the direction of the magnetic field. 

\begin{figure*}
\includegraphics[width=1.0\textwidth]{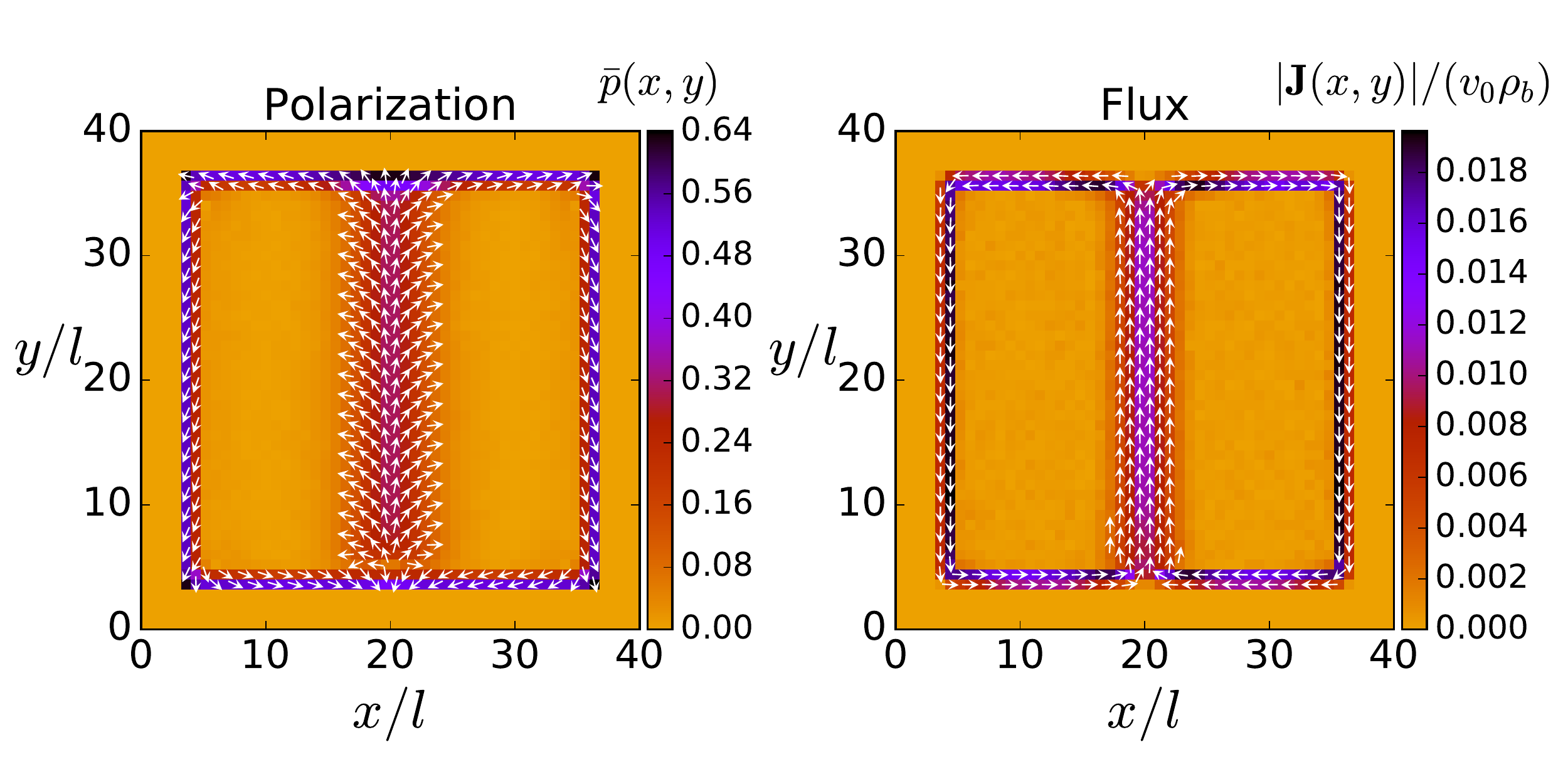}
\caption{Polarization and flux in a confined system. The results are obtained from Brownian dynamics simulations with $\ff=20$, $\kappa(x)=8 \sin(2 \pi x/5)$, $\gamma=1$, $D_r=20$
and $m=0.002$ in a system with walls at $x=0$, $y=0$, $x=10$ and $y=10$.
The self-propulsion speed of the ABPs is $v_0=\ff/\gamma$, $l$ is their persistence length and $\rho_b$ is the bulk density.
The walls are modeled by a WCA potential (see App. \ref{app:numerics}).
In the bulk the density, polarization and flux in the system are similar to those in a system without walls.
The polarization of ABPs near a wall is know to be directed towards the wall~\cite{speck2016ideal};
in this system, however, the orientation near the walls along the $y$ direction is directed downwards.
This is in line with the prediction if one considers the gradient of the magnetic field at those positions.
With the walls there are additional polarization effects:
there is also polarization parallel to the walls in the $x$ direction.
The polarization along the walls is related to the flux along the walls.
Polarization and flux along the walls have also been observed in the case of chiral particles \cite{caprini2019active}.
One can induce multiple 'lanes' of bulk flux with opposing direction by decreasing the period of the magnetic field. }\label{fig:walls}
\end{figure*}

In case of a finite system with walls, the bulk properties of the system are similar to that of an infinite system;
however, there is also flux parallel to the wall resulting in closed-loop fluxes (see Fig. \ref{fig:walls}).
A flux along walls has recently been investigated in the case of chiral active particle \cite{caprini2019active}.
The flux along the wall induced by Lorentz force can possibly be explained by the same mechanism.

\section{Active diffusion under Lorentz force}\label{sec:active_diff}
\begin{figure}[t]
\includegraphics[width=1.\columnwidth]{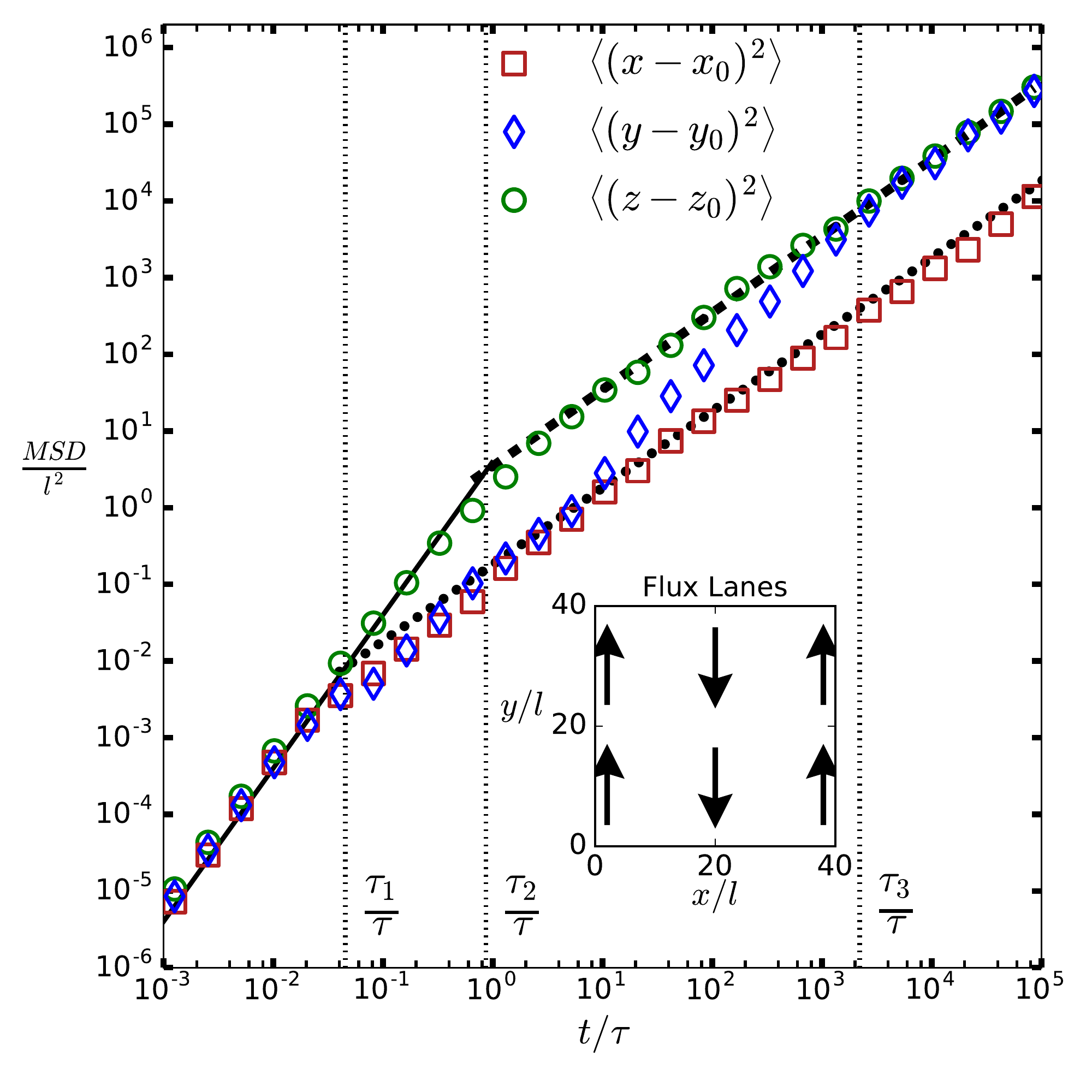}
\caption{The mean square displacement (MSD) of an ABP for a periodic system with $f=20$, $\kappa(x)=8 \sin(2\pi x/10)$, $\gamma=1$ and $D_r=20$.
$l$ is the persistence length of the ABP, and $\tau=1/2D_r$ its orientational correlation time.
Symbols represent results from Brownian dynamics simulations.
Diffusion parallel to the magnetic field orientation is unaffected by the field.
At long times, $\left<\Delta z(t) ^2\right> \sim 2D_z t$, shown as the dashed line, with $D_z$ the diffusion coefficient of a freely diffusing ABP [Eq. \eqref{Dz}].
The motion of an ABP along the $x$ is also diffusive at long times $\left<\Delta x(t) ^2\right> \sim 2D_x^{eff} t$ (dotted line) albeit with a reduced diffusion coefficient as given in Eq. \eqref{Dx}.
This is expected since a magnetic field hinders the motion of particle in the plane perpendicular to its orientation.
The motion along the $y$ direction, however, is affected differently than along $x$, which is related to the flux lanes (see inset).
At short times, $t < \tau_2$, the motion along $y$ is same as $x$.
The time scale of the transition from ballistic to diffusive motion in the $z$ direction is $\tau_2$ [Eq.~\eqref{tau2}].
However, at long times $t > \tau_3$, the rate of diffusion along $y$ becomes the same as that of a freely diffusing ABP.
$\tau_3$ is the time scale of the transition to diffusive motion in the $y$ direction [Eq.~\eqref{tau3}].
At intermediate times $\tau_2 < t < \tau_3$, the motion along $y$ is super diffusive. }\label{fig:msd}
\end{figure}

Here we investigate the mean squared displacement (MSD) of an ABP in the three directions.
The MSD in the $y$ direction is particularly interesting: for a magnetic field varying in the $x$ direction, the steady-state fluxes in the $y$ direction constitute layers of alternating flows (see inset to Fig.~\ref{fig:msd} and Fig.~\ref{fig:bulk_results}).
How does the emergence of these layers of alternating flows influence the MSD in the $y$ direction?

Free ABPs have a swim speed of $f/\gamma$ \cite{merlitz2017directional}; therefore, on short time scales the motion of ABPs is ballistic and is the same in the three directions:
\begin{align}
\left<\Delta x(t) ^2\right> = 
\left<\Delta y(t) ^2\right> = 
\left<\Delta z(t) ^2\right> = \left(\frac{f t}{\gamma}\right)^2,
\end{align}
where $\Delta x(t)$ is $x(t)-x(0)$, and similarly for $\Delta y(t)$  and $\Delta z(t)$.
The diffusion parallel to the magnetic field orientation is unaffected by the field. This implies that the 
diffusion coefficient in the $z$ direction is that of a freely diffusing ABP, which is
\begin{align} \label{Dz}
D_z = \frac{T+\Delta T}{\gamma}.
\end{align}
Since a magnetic field hinders the motion in the plane perpendicular to its direction, the diffusion in the $x$ and $y$ directions is reduced.  For an inhomogeneous magnetic field, the spatially varying diffusion coefficient in the $x$ and $y$ directions can be obtained from the symmetric part of the tensor in Eq.~\eqref{Gamma_inv} as
\begin{align}
D_x(x)=D_y(x) = \frac{T+\Delta T}{\gamma} \frac{1}{1+\kappa(x)^2}.
\end{align}
On a length scale larger than the period of the magnetic field, the space dependence of the diffusion constant becomes irrelevant, and one can approximate the effective diffusion constant for MSD in the $x$ direction by averaging over space:
\begin{align}\label{Dx}
D_{x}^{(eff)} = \frac{T+\Delta T}{\gamma}
\int_0^L dx \frac{\rho(x)}{1+\kappa(x)^2},
\end{align}
where $L$ is the period of the magnetic field.

There are three time scales in the system.
Under the assumption that the length scale of ballistic motion is much smaller than the period of the field, the shortest time scale, $\tau_1$, is the transition from ballistic to diffusive motion in the $x$ direction.
This time scale can be obtained from $(f \tau_1/\gamma)^2 = 2 D_{x}^{(eff)} \tau_1$, and is 
\begin{align}
\tau_1 = \frac{2 \gamma^2 D_{x}^{(eff)}}{f^2}.
\end{align}
The second time scale, $\tau_2$, is the transition from ballistic to diffusive motion in the $z$ direction, which can be calculated from $(f\tau_2/\gamma)^2 = 2 D_z \tau_2$ as
\begin{align}\label{tau2}
\tau_2 = \frac{2\gamma^2 D_z}{f^2}.
\end{align}
The largest time scale, $\tau_3$, is the transition to diffusive motion in the $y$ direction.
The diffusion in the $y$ direction is increased due to the flux lanes in the system.
Because the net flux in the system is zero, the motion in the $y$ direction becomes diffusive on the same time scale as the time it takes to diffuse from on lane to another.
This time scale can be calculated from
$2 D_x^{(eff)} \tau_3 = (L/2)^2$ as 
\begin{align}\label{tau3}
\tau_3 = \frac{L^2}{8 D_x^{(eff)}}.
\end{align}

The MSD in the $y$ direction becomes subballistic on the same time scale as  the $x$ direction, namely $\tau_1$; however the diffusive regime starts later at $\tau_3$.
This means that there is a super diffusive regime that persists over several decades in time.
Figure ~\ref{fig:msd} shows that when $t>\tau_3$ the MSD in the $y$ direction has the same effective diffusion coefficient as the $z$ direction.
The magnetic field hinders diffusion in both $x$ and $y$ direction.
Whereas the diffusion in $x$ direction is expectedly smaller than in the $z$ direction, the flux lanes increase the effective diffusion along $y$ and therefore the MSD in the $y$ direction 'catches up' with $z$ direction.

\section{Passive diffusion of tracer particle}\label{sec:passive_diff}

We showed in Sec~\ref{sec:noneq_ss} that the polarization [Eq.~\eqref{py}] is the fraction of the swim force that is converted into macroscopic motion. Since polarization scales as $l \kappa' \ll 1$, it is small in the limit where gradients are small compared to the persistence length of the ABP. To overcome this limitation, we consider a system in which the effect of the anisotropic motion of the ABPs on a tracer particle can accumulate.

\begin{figure}[t]
\vspace{-3cm}
\includegraphics[width=1.1\columnwidth]{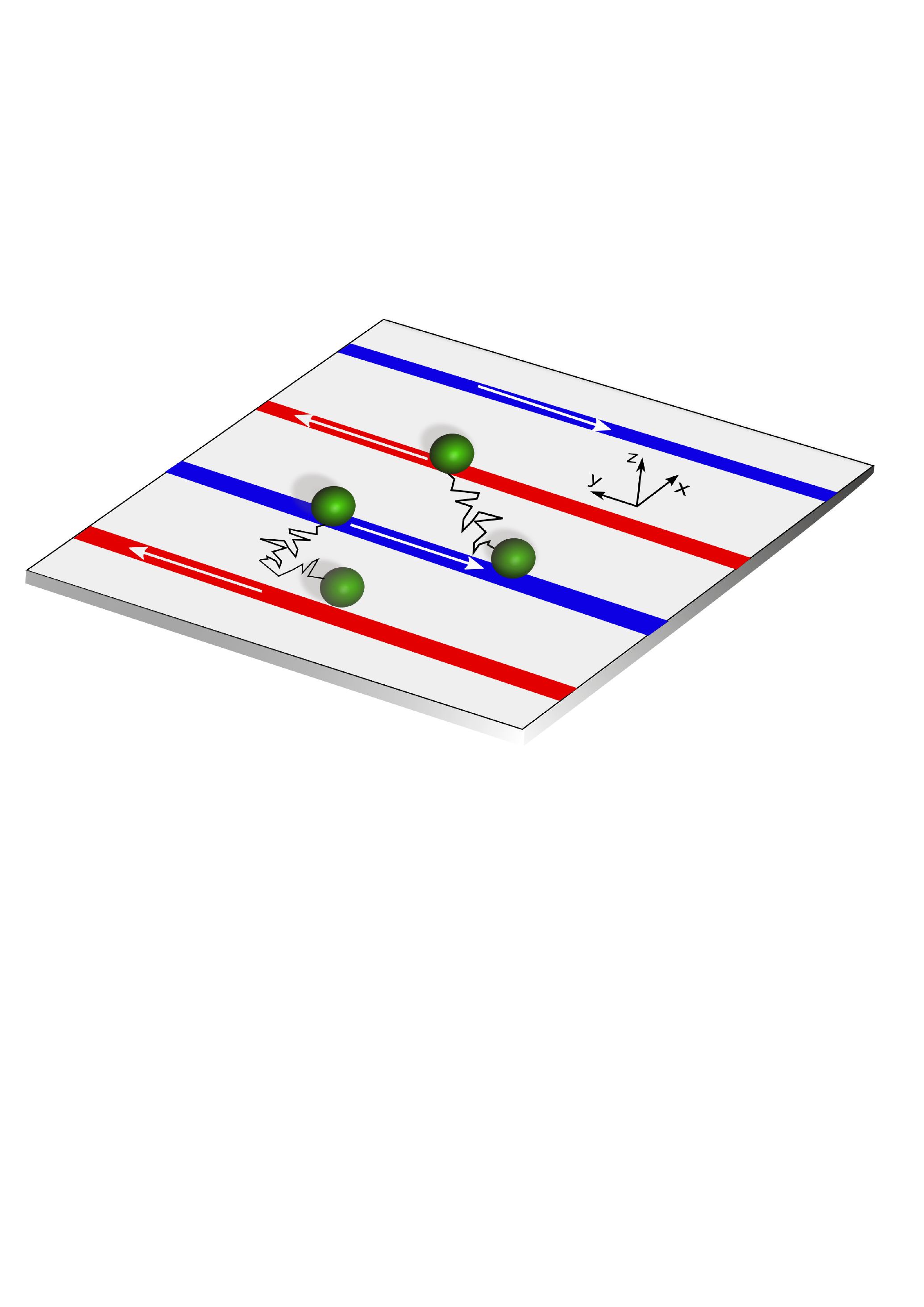}\vspace{-6cm}
\caption{Schematic of a passive tracer particle diffusing in the $xy$ plane.
The flux lanes are shown in blue and red.
The particle diffuses from one lane to the other shown as zigzag lines.
Inside a lane, the particle drifts along with the local velocity of the lane shown as white arrows.
This results in anisotropic diffusion of the particle.
Since the time spent by a particle inside a lane increases with decreasing diffusion coefficient, the anisotropy can be large for small diffusion coefficients.}\label{fig:tracer}
\end{figure}

Consider again the system of ABPs in steady state in a periodically varying magnetic field.
As discussed above the ABPs will form a periodic array of lanes carrying an alternating macroscopic flux in $y$ direction, see Fig.~\ref{fig:tracer}.
We now immerse an uncharged, passive tracer particle in the system and study its MSD.
On can expect that the steady-state fluxes of the ABPs affect the diffusion in the $y$ direction of the tracer particle.
Similar problems have been studied in different contexts, such as the diffusion of a passive particle in presence of random velocity fields~\cite{redner1989superdiffusive,redner1990superdiffusion, bouchaud1990superdiffusion} and the motion of polymer chains in random layered flows~\cite{sommer1996polymer}.

Due to the (alternating) velocity field, one can expect that the MSD of the tracer particle along $y$ direction is enhanced.
This implies that the motion of the particle is anisotropic.
The enhancement of the diffusion along the $y$ direction can be estimated as follows:
The particle diffuses freely with a diffusion rate $D_p$ between the velocity lanes.
Assuming that inside the lane, the particle moves with the local velocity, the distance that the particle travels along the lane is $l_f = v_y d ^2/2D_p$, where $v_y$ is the local velocity and $d$ is the width of the lane.
The particle experiences these flow lanes every $\tau_y = L^2/8D_p$ seconds, where $L/2$ is the distance between two neighboring lanes.
The increase in the diffusion rate along the $y$ direction can be estimated as  $ l_y^2/2\tau_y \sim v_y^2 d^4/D_pL^2$.
The anisotropy parameter, defined as the ratio of the MSDs in the $x$ and $y$ directions, is
\begin{align}\label{anisotropy_parameter}
r_a = 1 + \frac{v_y^2 d^4}{D_p^2L^2}.
\end{align}

An experimental measurement of the anisotropy of the diffusion of a tracer particle, as described above, would provide an, albeit indirect, experimental verification of the active fluxes in the system.
Since Lorentz force is relativistic in nature and ABPs are colloidal particles, one expects that $v_y$ is extremely small.
As we show below, this is indeed the case for the known experimentally realizable ABPs.
However, since the anisotropy parameter depends inversely on the diffusion coefficient of the tracer particle, it might still be observed experimentally for sufficiently large tracer particles on long time scales.
The smaller the diffusion constant of the tracer particle the longer it will stay in one lane and thus accumulating the effect by drifting along with the velocity field.
In the extreme case of a vanishing diffusion constant, $D_p$=0, the only observable displacement on a long
time scale is the drift and thus $r_a \gg 1$.

The anisotropy parameter can be estimated as follows:
The lane velocity can be obtained from Eq.~\eqref{yflux} as $v_y = j_y/\rho = v_0 \pb_y$, where $v_0 = f/\gamma$ is the self-propulsion speed and $\pb_y$ is the $y$ component of the polarization vector inside a lane.
It can be obtained from Eq.~\eqref{py} as $\pb_y \approx l q |\nabla B|/3\gamma$, where $l = v_0/2D_r$ is the persistence length and $|\nabla B|$ is the (maximum) gradient of the magnetic field. 
With this the lane velocity can be estimated as $v_y \sim v_0^2 q |\nabla B|/(2D_r \gamma)$.
To estimate the charge, we consider that the active particle is a Janus-shaped particle with half of its surface grafted with polymer chains, and the other half serves as the propulsion engine~\cite{jiang2010janus,walther2013janus,buttinoni2012active,samin2015self,wurger2015self}.
The total charge on such an ABP is $q=2\pi R^2 \sigma_b \lambda e$, where $R$ is the radius of the ABP, $\sigma_b$ is the grafting density, $\lambda$ is the number of charges per polymer chain, and $e$ is the elementary charge.
Using this together with the Stokes relations for the rotational diffusion coefficient $D_r$ and $\gamma$, one obtains
\begin{align}
v_y \approx \frac{4\pi}{3}\frac{v_0^2 \sigma_b \lambda e |\nabla B|}{T} R^4.
\end{align}

In order to estimate the magnitude of lane velocity, we consider $T = 298/k_b\,kgm^2s^{-2}$, where $k_b$ is the Boltzmann constant, a dense grafting density of $\sigma_b = 1 nm^{-2}$ and highly charged chains with $\lambda = 100$.
For the magnetic field we assume a period of $L \sim 1\,cm$.
For the magnetic field gradient we use $10\,T/cm$,
which can be obtained in magnetic matrices used for filtering fine and weakly magnetic particles~\cite{ge2017magnetic}.
For the ABPs we take the values $v_0 \approx 3 \,\mu m s^{-1}$ and $R \approx 2\,\mu m$ from Ref. \cite{lozano2016phototaxis}.
With these, we obtain the lane velocity $v_y \approx 2.5\times 10^{-3} \,\mu m s^{-1}$.
The width of a lane can be approximated as $d = L/(2\pi) \sim 0.1\,cm$.
Note that for the chosen parameters the diffusive Hall-constant is of the order, 
\begin{align}
\kappa = qB/\gamma \sim 0.1
\end{align}
implying that Lorentz force becomes comparable to frictional force on the particle.
Finally, assuming that the diffusion coefficient of the tracer particle is $D_p \sim 10^{-13}\, m^2/s$, a reasonable estimate for large tracer particles ($\sim \mu m$), we obtain from Eq. \eqref{anisotropy_parameter} $r_a \approx 2$.
Of course long-time scales are necessary to detect the effect.
However, the distance that a tracer particle drifts along a velocity lane in a day time turns out to be approximately $0.2\,mm$, which can be detected experimentally.
These estimates suggest that Lorentz forces may be observable experimentally for charged, active particles.
We note that it is not the magnitude of the magnetic field that is important for this Hall-like effect, but it is only the gradient which enters the calculation. 

A few remarks are necessary with respect to a possible experimental setup.
First, the total charge of the poly-electrolyte brush is compensated by counter-ions arising from the dissociation of the charged groups of polymers in water.
In a salt-free solution the counter-ions will be strongly localized inside the brush and thus effectively cancel the effect of the Lorentz-force~\cite{pincus1991colloid}.
However, a superior aspect of the external magnetic field is that it is not screened by electric charges.
Thus, at a high salt level in the solution the counter-ions are quasi-free \cite{borisov1994diagram,he2010polyelectrolyte}.
Furthermore, the exchange dynamics of them will be much faster than the Lorentz-induced drift.
Therefore, brush-decorated active colloidal particles in a salted solution could be a promising experimental system.

\section{Discussion and Conclusions}\label{sec:conclusion}
Lorentz force has the unique property that it depends on the velocity of the particle and is always perpendicular to it.
Although this force generates particle currents, they are purely rotational and do no work on the system.
As a consequence, the equilibrium properties of a Brownian system, for instance the steady-state density distribution, are independent of the applied magnetic field.
The dynamics, however, are affected by Lorentz force: the Smoluchowski equation picks up a tensorial (diffusion) coefficient, which reflects the anisotropy of the particle's motion~\cite{balakrishnan2008elements}.
The diffusion rate perpendicular to the direction of the magnetic field decreases with increasing field whereas the rate along the field remains unaffected.
In addition to this effect, Lorentz force gives rise to nondiffusive fluxes which are perpendicular to the density gradients~\cite{vuijk2019anomalous, chun2018emergence}.

The effects caused by the Lorentz force, however, occur only in nonequilibrium and cease to exist when the distribution of charged particles reaches equilibrium.
A system of ABPs, in contrast, is continuously driven out of equilibrium.
Lorentz force couples to the nonequilibrium dynamics of an ABP via its self-propulsion.
In this work, we showed that when subjected to an inhomogeneous magnetic field, a system of uniformly active ABPs settles into an unusual nonequilibrium steady state.
This state is characterized by (a) inhomogeneously distributed density with the same spatial symmetry as that of the magnetic field and (b) macroscopic fluxes perpendicular to the gradient of the magnetic field.

The theoretical approach is based on coarse graining of the Smoluchowski equation and yields analytical results for the density distribution, active fluxes, and polarization in the steady state.
We also showed that when combined with the linear-response theory, the theoretical predictions match the results from BD simulations.

That the density distribution in our system is inhomogeneous can be understood as follows.
The swim force increases the ABP's velocity in the direction of its orientation.
The Lorentz force rotates the velocity vector.
The effect of this rotation is a decrease in the speed of the particle in the direction of its orientation,
which is similar as decreasing the effect of the swim force.
From this it follows that a spatially varying magnetic field has a similar effect as ABP as subjecting them to an inhomgeneous activity field.
However, the existence of nondiffusive fluxes perpendicular to the density gradient in the steady-state is a unique feature of this system and is a consequence of the antisymmetric coupling of the $x$ and $y$ directions due to the Lorentz force.
In case of a finite system with walls, the bulk properties of the system are similar to that of an infinite system;
however, there is also flux parallel to the wall resulting in closed-loop fluxes 

The MSD of an ABP in the presence of an inhomogeneous magnetic field has nontrivial behavior related to the existence of flux lanes in the system.
Whereas the motion along the direction of the field is unaffected by the field, it is hindered in the plane perpendicular to the field.
The diffusion coefficient in the $z$ direction is, as expected, the same as that of a freely diffusing ABP.
The diffusion in the $x$ direction is expectedly smaller than in the $z$ direction; however, the same does not hold for the $y$ direction.
The diffusion in the $y$ direction is super diffusive at intermediate times.
At long times, the motion along the $y$ direction becomes diffusive with the same rate as along the $z$ coordinate.
The emergence of flux lanes in the $y$ direction gives rise to this nontrivial behavior of the MSD in the $y$ direction. 

Our study naturally raises the question whether the effects induced by Lorentz force in a system of ABPs can be observed experimentally.
A Lorentz force on a colloidal particle is extremely small due to the particle's low velocity.
However, for such a particle to be affected by a Lorentz force, it is not the absolute value of the Lorentz force that is important but how it compares with the frictional forces.
Since both frictional and Lorentz force depend on the velocity, for Lorentz force to have any significant effect on the motion of the particle the diffusive Hall constant $\kappa = qB / \gamma$ must be of the order of unity.
In Sec. \ref{sec:passive_diff}, we considered the possibility of experimental verification of this study where we have proposed to use the high total charge of a poly-electrolyte brush grafted to an ABP to obtain a value of $\kappa \simeq 0.1$.
Direct experimental verification of the flux lanes might still be difficult.
A possibility is to measure the effect of the flux lanes on the diffusive behavior of a passive tracer particle on which the effect of the flux lanes can accumulate. 
Using values for propulsion speed and particle size from experiments \cite{lozano2016phototaxis} and large but reasonable estimates for other parameters in the system, we estimated a drift of the  tracer particle along the flux lines of the active system of the order of millimetre per day.
This indicates that magnetic-field induced effect on a charged active system could be measured experimentally.

Finally, note some general aspects of the application of magnetic fields to soft-matter systems.
First of all, magnetic fields are not screened by static charges such as electric field are.
Strong electric field strength in particular combined with thin-film geometries are prone to short-cuts, irreversibly damaging the system. 
By contrast the application even of high magnetic fields is possible and is used in, for example, high-resolution NMR-experiments to medicine.
In this work we have shown that Lorentz forces may be detectable in active soft-matter systems, which opens new possibilities to manipulate such systems.
Interestingly, macroscopic effects are induced by the symmetry breaking of the magnetic field only and do not require a macroscopic flux of the particles which establishes a novel quality of the Hall-like effect, which in this case depends on the gradient of the field only.    



\appendix

\section{small-mass limit}\label{app:small_mass_limit}

In order to obtain the small-mass limit of  the FPE corresponding to Eqs.
 \eqref{sde_r}\eqref{sde_v} and \eqref{sde_p}, one can use the method described
in Chap. 10 of Ref. \cite{risken1989fokker}.
This was done for a single passive particle in a homogeneous
magnetic field in Ref. \cite{chun2018emergence}.
We start with the FPE corresponding to Eqs. \eqref{sde_r}, \eqref{sde_v} and \eqref{sde_p} for the time evolution of the probability density $P(t) \equiv P(\rr,\vv,\pp,t)$ is \cite{gardiner2009stochastic}
\begin{equation}\label{sml_FPE_P}
 \frac{\partial}{\partial t} P(t) = \left( L_{rev} + L_{irr} + L_{r} + L_{a} \right) P(t),
\end{equation}
where the time-evolution operator has been split up in a reversible part
\begin{align}
 L_{rev} P(t)=  -\vv &\cdot \nabla_{\rr}P(t)  \nonumber\\
 &+ \frac{1}{m} B(\rr) \nabla_{\vv} \cdot \left[ M \cdot \vv P(t) \right], 
\end{align}
an irreversible part without the  part coming from Eq. \eqref{sde_p}
\begin{align}
 L_{irr} P(t) =  \frac{\gamma}{m} \nabla_{\vv} \cdot \left[ \vv P(t) + \frac{T}{m} \nabla_{\vv} P(t) \right],
\end{align}
the part representing the rotation of the orientation vector
\begin{align}
 L_{r} P(t) =  D_r \sR^2 P(t),
\end{align}
with the rotation operator $\sR=\pp\times\nabla_{\pp}$ \cite{morse1953methods},
and the active, nonequilibrium operator coming from the self-propulsion
\begin{align}
 L_{a} P(t) = - \frac{\ff}{m} \pp \cdot \nabla_{\vv} P(t).
\end{align}
Equation \eqref{sml_FPE_P}  is equivalent to
\begin{equation}\label{sml_FPE_P_transform}
 \frac{\partial}{\partial t} \bar P(t)=\left( \bar L_{rev}+ \bar L_{irr}+ \bar L_{r}+ \bar L_{a} \right) 
 \bar P(t),
\end{equation}
where
\begin{equation}\label{sml_P_transform}
 \bar{P}(t)=P(t)R(\vv)^{-1/2}
\end{equation}
and
\begin{equation}
 \bar{L} = R(\vv)^{-1/2} L R(\vv)^{1/2},
\end{equation}
where $L$ can be any of the operators in Eq. \eqref{sml_FPE_P}, and 
\begin{equation}
 R(\vv)=\left( \frac{m}{2 \pi T} \right)^{3/2} e^{-\frac{m}{2T} \vv^2}
\end{equation}
is the solution to $L_{irr} R(\vv) =0$, normalized such that the integral over $\vv$ is one.
The transformed operators are
\begin{align}
  \bar L_{rev} = - \sqrt{\frac{T}{m}} & \nabla_{\rr} \cdot \left( \bb^\dagger + \bb \right) 
+ \frac{q}{m} \B(\rr) \cdot \left( \bb^\dagger \times \bb \right),
 \end{align}
\begin{equation}
 \bar L_{irr} = -\frac{\gamma}{m}\bb^\dagger \cdot \bb,
\end{equation}

 \begin{align}
  \bar L_{r} = L_{r}
 \end{align}
 and
 \begin{align}
  \bar L_{a} =  \frac{\ff}{\sqrt{Tm}} \pp \cdot \bb^\dagger,
 \end{align}
 where $\bb = \sqrt{T/m} \nabla_{\vv} + \frac{1}{2}\sqrt{m/T} \vv $ and
 $\bb^\dagger = - \sqrt{T/m} \nabla_{\vv} + \frac{1}{2}\sqrt{m/T} \vv $.

 The eigenfunctions of the operator $b^{\dagger}_{\alpha} b_\alpha$, where $\alpha$ is $x$,$y$ or $z$, are
 \begin{align}
 \psi_0(v_\alpha) = \left( \frac{m}{2 \pi T} \right)^{1/4} e^{-\frac{m}{4T} v_\alpha^2},
 \end{align}
 and
 \begin{align}
  \psi_n(v_\alpha) =  \frac{\psi_0(v_\alpha)}{\sqrt{n!2^n}} H_n\left( \sqrt{\frac{m}{2T}} v_\alpha \right) ~\text{for}~ n>0,
 \end{align}
where  $H_n$ are Hermite polynomials.
The operators $b_\alpha^\dagger$ and $b_\alpha$ are the raising and lowering operators of the eigenfunctions:
$b_\alpha^\dagger \psi_n(v_\alpha)=\sqrt{n+1} \psi_{n+1}(v_\alpha)$ and
$b_\alpha \psi_n(v_\alpha)=\sqrt{n} \psi_{n-1}(v_\alpha)$.
Because the eigenfunctions  are orthonormal,
\begin{equation}
 \int_{-\infty}^{\infty} dx \psi_n(x) \psi_m(x) = \delta_{n,m},
\end{equation}
 they can be used to expand $\bar P(t)$:
\begin{equation}\label{sml_P_expansion}
 \bar P(t) = \sum_{n_x,n_y,n_z = 0}^\infty c_{n_x,n_y,n_z} \psi_{n_x}(v_x) \psi_{n_y}(v_y) \psi_{n_z}(v_z),
\end{equation}
where $c_{n_x,n_y,n_z} \equiv c_{n_x,n_y,n_z}(\rr,\pp,t)$.

The probability density for the position and orientation, $Q(t)\equiv Q(\rr,\pp,t)$, is given by the 
first expansion coefficient:
\begin{align}
 Q(t) =& \int d \vv P(t) \\
 =& \int d \vv \bar P(t) \psi_0(v_x)\psi_0(v_y)\psi_0(v_z) \\
 =& c_{0,0,0}.
\label{sml_Q_c0}
\end{align}
The translational flux, 
\begin{align}
 \J(\rr,\pp,t) &= \int d\vv ~ \vv P(\rr,\vv,\pp,t),
\end{align}
can be evaluated by using Eqs. \eqref{sml_P_transform}, \eqref{sml_P_expansion} and 
$v_{\alpha} \psi_0(v_\alpha) = \sqrt{\frac{T}{m}} \psi_1(v_\alpha)$:
\begin{align}\label{sml_translational_flux}
\J(\rr,\pp,t) = \sqrt{T/m} \cc_1(\rr,t).
\end{align}
Using this, together with the definition of the flux in orientation space
\begin{align}\label{sml_rotational_flux}
\J_{rot}(\rr,\pp,t) = - D_r \sR Q(t),
\end{align}
one obtains the following equation for the time evolution of the probability density:
\begin{align}\label{sml_FPE_Q_flux}
\partial_t Q(t) = -\nabla_{\rr} \cdot \J - \sR \cdot \J_{rot}.
\end{align}

Next, we need an expression for the $\cc_1$ in terms of $Q(t)$.
We are interested in the case where the magnetic field is oriented perpendicular to the gradient of the field,
and take this to be the $z$ direction,
so $\B(\rr) = B(\rr) \hat \zz$.
Equation \eqref{sml_FPE_P_transform} together with the orthonormality of the eigenfunctions yields an hierarchy of equations 
for the functions $c_{n_x,n_y,n_z}$ called a Brinkman hierarchy \cite{brinkman1956brownian}:
\begin{widetext}
 \begin{align}\label{sml_brinkman}
  \frac{\partial}{\partial t} c_{n_x,n_y,n_z} =& -\frac{\gamma}{m} c_{n_x,n_y,n_z}(n_x+n_y+n_z) - \D \cdot 
    \begin{bmatrix}
     \sqrt{n_x+1} c_{n_x+1,n_y,n_z}\\
     \sqrt{n_y+1} c_{n_x,n_y+1,n_z}\\
     \sqrt{n_z+1} c_{n_x,n_y,n_z+1}
    \end{bmatrix}
    -\hat \D \cdot
    \begin{bmatrix}
     \sqrt{n_x} c_{n_x-1,n_y,n_z}\\
     \sqrt{n_y} c_{n_x,n_y-1,n_z}\\
     \sqrt{n_z} c_{n_x,n_y,n_z-1}
    \end{bmatrix} \nonumber \\
   & + \frac{qB(\rr)}{m} \sqrt{n_x(n_y+1)} c_{n_x-1,n_y+1,n_z}
    - \frac{qB(\rr)}{m} \sqrt{(n_x+1)n_y} c_{n_x+1,n_y-1,n_z}
    + D_r \sR^2 c_{n_x,n_y,n_z},
 \end{align}
\end{widetext}
where $\D = \sqrt{\frac{T}{m}} \nabla_{\rr}$ and
$\hat \D = \sqrt{\frac{T}{m}} \nabla_{\rr} - \frac{1}{\sqrt{Tm}} f \pp $.

The equation governing the time evolution of the first expansion coefficient is
\begin{align}\label{sml_c0}
 \frac{\partial}{\partial t} c_{0,0,0} = - \D \cdot \cc_1  + D_r \sR^2 c_{0,0,0},
 \end{align}
where $\cc_1 = (c_{1,0,0}, c_{0,1,0}, c_{0,0,1})^T$.
This is of course the same as Eq. \eqref{sml_FPE_Q_flux}.
The equation for $\cc_1(t)$ is again a differential equation in $t$, which can be solved after a Laplace transformation:
\begin{align}\label{sml_c1}
\widetilde\Gamma_m^{(1)}(s) \cdot \widetilde{\cc}_1(s) = \frac{m}{\gamma}
& [ \cc_1(0) + D_r \sR^2 \widetilde{\cc}_1(s) \nonumber \\
& - D_{1,2} \cdot \widetilde{\cc}_2(s) - \hat \D \cdot \widetilde{c}_{0,0,0}(s) ],
\end{align}
where the tilde indicates a Laplace transformation from the variable $t$ to $s$, $D_{1,2}$ is a matrix of which the entries are linearly proportional to $D_x$, $D_y$ or $D_z$,
$\cc_2(t)$ is a vector of which the elements are all $c_{n_x,n_y,n_z}$ such that $n_x+n_y+n_z=2$, and
\begin{align}
\widetilde\Gamma_m^{(1)}(s) = 
    \begin{bmatrix}
	1+s\frac{m}{\gamma} & -\kappa(\rr)           & 0 \\
	\kappa(\rr)           & 1+s\frac{m}{\gamma} & 0 \\
	0                   & 0     & 1+s\frac{m}{\gamma}
    \end{bmatrix}.
\end{align}
Note that $\widetilde\Gamma_m^{(1)}(s) =\Gamma(\rr) + \mathcal{O}(m) $.
The Laplace transformation of the equation for $\cc_2(t)$ is
\begin{align}\label{sml_c2}
\Gamma_m^{(2)}(\rr,s) \cdot \widetilde{\cc}_2(s) = \frac{m}{\gamma}& [ \cc_2(0) + D_r \sR^2 \widetilde{\cc}_2 
    \nonumber \\
& -D_{2,3} \cdot \widetilde{\cc}_3 - \hat D_{2,1} \cdot \widetilde{\cc}_1
],
\end{align}
where $\widetilde\Gamma_m^{(2)}$ is a matrix of which the entries are at least of order $m^0$, $D_{2,3}$ is a matrix whose elements are proportional to $D_x$, $D_y$ or $D_z$, $\hat D_{2,1}$ is a matrix whose elements are proportional to $\hat D_x$, $\hat D_y$ or $\hat D_z$, and 
$\cc_3(t)$ is a vector of which the elements are all $c_{n_x,n_y,n_z}$ such that $n_x+n_y+n_z=3$.

By only retaining the zeroth-order contribution in $m$ to Eq. \eqref{sml_c0}, one obtains an equation that is independent of $v$ and holds for $s m/\gamma << 1$ (that is, for times longer than the velocity autocorrelation time).
The vector $\D$ is of order $m^{-1/2}$, so only the terms of order $m^{1/2}$ of $\cc_1$ are needed, which means that one does not need the matrices $D_{1,2}$, $D_{2,3}$ and $\hat D_{2,1}$.
Equation \eqref{sml_c2} can be used to replace the $\cc_2$ in Eq. \eqref{sml_c1}, which, after inverting the Laplace transformation and only retaining terms of order $m^{1/2}$ or lower, becomes
\begin{align}
\Gamma \cdot \cc_1(t) = - \frac{m}{\gamma} \hat D c_{0,0,0} + \mathcal{O}(m),
\end{align}
which gives for the translational flux
\begin{align}\label{sml_translational_flux_full}
\J(\rr,\pp,t) =-\frac{1}{\gamma} \Gamma^{-1}\cdot (T \nabla_{\rr} - f \pp)Q(t) + \mathcal{O}(m^{1/2}),
\end{align}
where
\begin{align}
\Gamma^{-1}(\rr) = \boldsymbol{1} - \frac{\kappa(\rr)}{1+ \kappa^2(\rr)} M
+  \frac{ \kappa^2(\rr)}{1+ \kappa^2(\rr)} M^2.
\end{align}

The result for the translational flux together with Eqs. \eqref{sml_Q_c0} and \eqref{sml_c0} gives following Fokker-Planck equation:
\begin{align}\label{sml_FPE_Q_1}
\partial_t Q(t) =  \frac{1}{\gamma}\nabla \cdot \left[ \Gamma^{-1} \cdot ( T \nabla_{\rr} - f\pp ) Q(t) \right]
- D_r \sR^2 Q(t).
\end{align}
%
%
%

\section{Gradient expansion}\label{app_gradient_expansion}

The goal of the gradient expansion is to integrate out the orientational degrees of freedom of the probability density $Q(\rr,\pp,t)$ and obtain a drift-diffusion equation for the probability density $\rho(\rr,t)$ of the position variable \cite{tailleur2008statistical,cates2013when}.
This expansion is based on a decomposition of the probability density $Q(\rr,\pp,t)$ in spherical harmonics:
\begin{align}\label{ge_Q_expansion}
Q(\rr,\pp,t) = \rho + \ssig \cdot \pp + 
\ttau: \left(\pp \pp - \boldsymbol{1}/3\right) +
\Omega,
\end{align}
where the scalar $\rho$, the vector $\ssig$ and the matrix $\ttau$ are function of $\rr$ and $t$.
These functions are linear combinations of, respectively, the zeroth, first and second spherical harmonic components of $Q$, and $\Omega$ is the projection onto the third and higher-order spherical harmonics.
Using this decomposition in Eqs. 
\eqref{sml_rotational_flux}, \eqref{sml_FPE_Q_flux} and 
\eqref{sml_translational_flux_full}
and integrating over the orientational degrees of freedom gives
\begin{align}\label{ge_FPE_rho}
\frac{\partial}{\partial t}\rho = - \nabla_{\rr} \cdot \jj,
\end{align}
where
\begin{align}\label{ge_j1}
\jj(\rr,t) = - \frac{1}{\gamma}\Gamma^{-1} \cdot \left[
T \nabla_{\rr} \rho - \frac{\ff}{3} \ssig \right]
\end{align}
is the flux in position space.
An equation for $\ssig$ can be obtained by multiplying Eq.
\eqref{sml_FPE_Q_flux} by $\pp$ and then integrating over the orientational degrees of freedom.
This gives
\begin{align}\label{ge_sig1}
\frac{\partial}{\partial t} \sigma_a = 
&\frac{T}{\gamma} \partial_b \left( \Gamma^{-1}_{b c} \partial_c \sigma_a \right) - \nonumber \\
&\frac{\ff}{\gamma} \partial_b \left( \frac{2}{5} \Gamma^{-1}_{b c}  \tau_{c a} + \Gamma^{-1}_{b a} \rho \right) - 2 D_r \sigma_a,
\end{align}
where $a$, $b$ and $c$ are the indices of the vectors and matrices, and $\Gamma^{-1}_{a b}$ denotes the $a$ $b$ component of the matrix $\Gamma^{-1}$.
Similarly, an equation for $\ttau$ can be obtained by multiplying Eq.
\eqref{sml_FPE_Q_flux} by $\pp\pp-\boldsymbol{1}/3$ and then integrating over the orientational degrees of freedom:
\begin{align}\label{ge_tau1}
\frac{\partial}{\partial t} \tau_{ab} =& 
\frac{T}{\gamma} \partial_c \left(\Gamma^{-1}_{cd} \partial_d \tau_{ab} \right) - \nonumber \\
&\frac{\ff}{2\gamma} \partial_c \left( \Gamma^{-1}_{ca} \sigma_b + \Gamma^{-1}_{cb} \sigma_a - \frac{2}{3} \Gamma^{-1}_{cd} \delta_{ab} \right) - \nonumber \\
& 6 D_r \tau_{ab} + \partial_c \Upsilon_{abc},
\end{align}
where the $\partial_c \Upsilon_{cab}$ comes from the projection on to the third and higher-order harmonics.
Equations \eqref{ge_FPE_rho} \eqref{ge_sig1} \eqref{ge_tau1} are exact, but not closed since $\Upsilon$ is not specified.

We now consider the limit in which the gradients in the system are small in comparison to the persistence length of the ABP~\cite{schnitzer1993theory,cates2013when}.
The relaxation time of $\rho$ is of order $\sim (\nabla)^{-1}$, whereas all the other harmonics relax in times of order $\sim 1$.
This implies that in the limit of small gradients, $\rho$ is the slowest mode.
We thus assume that the time derivative of $\ssig$ and
terms with both a $\ssig$ and  a $\nabla$ can be neglected. This yields
\begin{align}\label{ge_sigma}
\ssig= - \frac{\ff}{2 D_r \gamma}  \nabla_{\rr} \cdot \left[ \Gamma^{-1} \rho \right],
\end{align}
 which is related to the polarization by $\ppb = \frac{1}{3} \ssig/ \rho$.
With this expansion the flux becomes
\begin{align}\label{ge_j}
\jj(\rr,t) = - \frac{\ff^2}{6 D_r \gamma^2} \Gamma^{-1} \cdot
	 \nabla \cdot \left( \Gamma^{-1} \rho \right)  & - \frac{T}{\gamma} \Gamma^{-1} \cdot \nabla \rho.
\end{align}

In a system in steady state with periodic boundary conditions and a magnetic field of the form $B(x)\hat{e}_z$,
the flux in the x direction must be zero (see Sec. \ref{sec:noneq_ss}).
This condition together with Eq. \eqref{ge_j} yields the steady-state density:
\begin{align}\label{ge_rho}
\rho(x) \propto \left[ 1 + \kappa(x)^2 \right]^{\delta/2},
\end{align}
where the proportionality constant is determined by the normalization of the density, $\delta=\frac{\Delta T}{T+\Delta T}$ and $\Delta T = \frac{\ff^2}{6 D_r \gamma}$.
Using this density to calculate the $y$ component of the steady-state flux gives
\begin{align}\label{ge_yflux}
j_y(x) = \frac{\ff}{\gamma}  \pb_y(x)\rho(x).
\end{align}
The steady-state polarization in the $x$ and $y$ directions can be determined from Eq. \eqref{ge_sigma} and the density:
\begin{align}\label{ge_px}
\pb_x = - \frac{(\delta - 2) l}{3}
    \frac{\kappa(x) \kappa'(x)}{\left[1+\kappa(x)^2 \right]^2}
\end{align}
\begin{align}\label{ge_py}
\pb_y = - \frac{l}{3}
    \frac{\kappa'(x)}{\left[1+\kappa(x)^2 \right]^2}
    \left[ 1+  (1-\delta)\kappa(x)^2 \right]
\end{align}

\section{Linear-response theory}\label{app:linear_response}
Linear-response theory for ABPs has been developed in Ref. \cite{sharma2016green-kubo} to study the average swim speed,
and was later applied to a system of interacting ABPs with inhomogeneous activity to study the density profile \cite{sharma2017brownian} and flux \cite{merlitz2018linear}.
Here we adapt the linear-response theory for ABPs and apply it to an charged ABP in a space-dependent magnetic field.
\begin{equation}\label{lr_FPE_Q}
 \frac{\partial}{\partial t} Q(t) = \left( \mathcal{L}_{d} + 
 \mathcal{L}_{r} + \mathcal{L}_{a} \right) Q(t),
\end{equation}
where
\begin{equation}
\mathcal{L}_{d} Q(t) =  \frac{T}{\gamma} \nabla_{\rr} \cdot
\left[  \Gamma^{-1} \cdot \nabla_{\rr}  Q(t) \right]
\end{equation}
is the operator corresponding to the particle's diffusive behavior,
$\mathcal{L}_{r}=L_{r}$ is the rotational time-evolution operator and 
\begin{equation}
\mathcal{L}_{a} Q(t) = -  \frac{\ff}{\gamma} \nabla_{\rr} \cdot \left[ \Gamma^{-1} \cdot \pp Q(t) \right]
\end{equation}
is the nonequilibrium time-evolution operator.
Next, we split the probability density $Q(t)$ in an equilibrium ($Q_{eq}$) 
and a nonequilibrium ($\delta Q(t)$) part,
where the equilibrium part is the steady-state solution to Eq. \eqref{lr_FPE_Q} with $\ff=0$, 
which is constant in space.
The equation for the nonequilibrium part is
\begin{equation}
 \frac{\partial}{\partial t} \delta Q(t) = \left( \mathcal{L}_{d} + \mathcal{L}_{r}
 + \mathcal{L}_{a} \right) \delta Q(t) + \mathcal{L}_{a} Q_{eq},
\end{equation}
because $\partial_t Q_{eq} = 
 \left(  \mathcal{L}_{d} + \mathcal{L}_{r} \right) Q_{eq} = 0$.

We are interested in nonequilibrium steady state averages,
 and therefore assume that the system started out at $t=-\infty$ in the equilibrium state.
The previous equation can be integrated from $t=-\infty$ to $t=0$ by treating $\mathcal{L}_{a} Q_{eq}$ as an inhomogeneity:
\begin{equation}\label{lr_deltaQ}
 \delta Q_{ss} =  - \frac{\ff}{\gamma} \int_0^\infty d\tau e^{\tau \left( \mathcal{L}_{d}  + \mathcal{L}_{r}
 + \mathcal{L}_{a} \right)} Q_{eq}   \left[ \nabla_{\rr} \cdot \Gamma^{-1}(\rr) \right] \cdot \pp,
\end{equation}
where $\delta Q_{ss} = \delta Q(t=\infty)$ is the steady-state solution of the nonequilibrium part of the probability density.
This equation is exact and can be used to calculate the steady-state nonequilibrium average of a 
generic function  $g=g(\rr,\pp)$:
\begin{align}
 \left< g \right>^{\rr,\pp} &= \int d\rr \int d\pp ~ Q_{ss}(\rr,\pp) g \\
&= \left< g \right>^{\rr,\pp}_{eq} \nonumber \\
& - \frac{f}{\gamma} \int_0^\infty  dt \left< 
\left[ \nabla_{\rr} \cdot \Gamma^{-1}(\rr) \right] \cdot \pp e^{t \left( \mathcal{L}_{d}^\dagger +
\mathcal{L}_{r}^\dagger + \mathcal{L}_{a}^\dagger\right) } g \right>^{\rr,\pp}_{eq},
\end{align}
where $Q_{ss}(\rr,\pp) = Q_{eq} + \delta Q_{ss}(\rr,\pp)$ is the nonequilibrium steady-state probability density, 
the super script $\rr,\pp$ denotes the variables that are averaged over, the subscript $eq$ indicates that the average is taken with respect to the equilibrium distribution, 
and a $\dagger$ indicates the adjoint of an operator.
 The adjoint operators acting on a function $g$ are 
 \begin{equation}
 \mathcal{L}_{d}^\dagger g(\rr,\pp) =   \frac{T}{\gamma} \nabla_{\rr}  \cdot \left( \Gamma^{-1,T} \cdot \nabla_{\rr} g \right)  ,
\end{equation}
 \begin{equation}
  \mathcal{L}_{r}^\dagger g(\rr,\pp) =   D_r  \nabla_{\pp} \times \left[ \pp \cdot  \left( \nabla_{\pp} \times \left(\pp g \right) \right) \right],
 \end{equation}
 and
 \begin{equation}
  \mathcal{L}_{a}^\dagger g(\rr,\pp) =  \frac{\ff}{\gamma} \pp \cdot \left( \Gamma^{-1,T} \cdot \nabla_{\rr} g \right),
 \end{equation}
where the superscript $T$ indicates a transpose.

The response of the system to the activity up to linear-order in $\ff$
is obtained by removing from the 
full time-evolution operator the nonequilibrium operator $\mathcal{L}_{a}^\dagger$:
\begin{align}\label{lr_average}
 \left<g \right>^{\rr,\pp}_{lr} =& \left<f\right>^{\rr,\pp}_{eq}\nonumber \\
& - \frac{\ff}{\gamma} \int_{0}^{\infty}  dt \left< 
\left[ \nabla_{\rr} \cdot \Gamma^{-1}(\rr) \right] \cdot \pp e^{t \left( \mathcal{L}_{d}^\dagger +
\mathcal{L}_{r}^\dagger \right) } g \right>^{\rr,\pp}_{eq}.
\end{align}
This Green-Kubo relation defines the active transport coefficient $\alpha$ corresponding to an functions $g$ by $\alpha = \lim_{f\to 0} \left( <g>_{lr} - <g>_{eq}\right)/f$.
Equation \eqref{lr_average} shows that only quantities that are odd functions of $\pp$ have a nonzero linear response
because the angular average in equilibrium  any function of an odd power of $\pp$  yields zero by symmetry.

\section{Polarization from linear-response theory}\label{app:orientation}
Here we calculate the polarization using the general linear-response theory developed in the previous appendix.
The steady-state polarization is defined as the average orientation per particle:
\begin{equation}\label{or_average_p}
 \ppb(\rr') = \frac{\left<  \pp \delta\left(\rr' - \rr \right) \right>^{\rr,\pp}}{\rho(\rr')},
\end{equation}
where $\rho(\rr')$ is the average steady-state density, which is the same as the steady-state probability density of particles as a function of the position alone.
Because the  density operator $\hat{\rho}(\rr') = \delta(\rr'-\rr)$ is independent of the orientation, the density has 
no linear response to the activity; therefore,
when one only considers contributions up to linear order in the swim force to the polarization, the position-dependent density can be replaced by the bulk density $\rho_b$.
 
The orientation up to linear order in $\ff$ can be calculated using Eq. \eqref{lr_average} to evaluate
the average in Eq. \eqref{or_average_p}:
\begin{align}\label{or_plin_full1}
 \ppb^{(1)}(\rr') = &- \frac{\ff}{\rho_b\gamma} \int_0^\infty dt \nonumber \\
 & \left<  \left( \nabla_{\rr} \cdot
 \Gamma^{-1} \right) \cdot \pp e^{t \left( \mathcal{L}_{d}^\dagger
 + \mathcal{L}_{r}^\dagger \right)}
 \pp \delta(\rr'-\rr) \right>^{\rr,\pp}_{eq},
\end{align}
where the superscript $(1)$ indicates that it is correct up to first order in the self-propulsion force $\ff$.
Because in equilibrium the position is not correlated with the orientation,
one can integrate out the orientational degrees of freedom in the average.
The orientational integral in the equilibrium average in Eq. \eqref{or_plin_full1} is
\begin{equation}
 \left< \pp e^{t \mathcal{L}^\dagger_{r}} \pp \right>^{\pp}_{eq} = \left<\pp(0)\pp(t) \right>,
\end{equation}
where the $\pp(t)$ is the solution of the stochastic process resulting from Eq. \eqref{sde_p} with initial orientation $\pp(0)$, and the angle brackets indicate an average over realizations of the noise.
This is the autocorrelation function for Brownian motion on a unit sphere and can be calculated by expanding
the corresponding FPE in spherical harmonics  \cite{farage2015effective} as
\begin{equation}
 \left<  \pp(t) \pp(t') \right> = \frac{1}{3} e^{-2D_r|t-t'|} \boldsymbol{1}.
\end{equation}

With the result for the autocorrelation of $\pp(t)$, Eq. \eqref{or_plin_full1} becomes
\begin{align}\label{or_plin_full}
 \ppb^{(1)}(\rr') = - \ff & \int_0^\infty  dt  \frac{e^{-2D_r t}}{3 \rho_b \gamma}  \nonumber \\
 & \left<  \left( \nabla_{\rr} \cdot
 \Gamma^{-1} \right) e^{t \mathcal{L}_{d}^\dagger }
  \delta(\rr'-\rr) \right>^{\rr}_{eq},
\end{align}
where average is taken with respect to the equilibrium probability density for the position
variable.
By adding an other integral, one can take the gradient of $\Gamma^{-1}$ out of the average:
\begin{align}\label{or_plin_r''}
 \ppb^{(1)}(\rr') &= - \ff \int_0^\infty dt  \frac{e^{-2D_r t}}{3 \rho_b \gamma} \int d\rr''  \nonumber \\
 &\left[ \nabla_{\rr''} \cdot \Gamma^{-1}(\rr'') \right] 
  \left<  \delta(\rr''-\rr) e^{t \mathcal{L}_{d}^\dagger  }
  \delta(\rr'-\rr) \right>^{\rr}_{eq},
\end{align}
which can be written as
\begin{equation}
 \ppb^{(1)}(\rr') = \ff \int_0^\infty dt\int d\rr'' ~ \boldsymbol{\chi}\left( \rr',\rr'',t\right),
\end{equation}
where the response function $\boldsymbol{\chi}\left( \rr',\rr'',t\right)$ is
\begin{equation}
 \boldsymbol{\chi}\left( \rr',\rr'',t\right) =-\frac{1}{3\gamma}e^{-2 D_rt}
 G(\rr',\rr'',t) \nabla_{\rr''} \cdot \Gamma^{-1}(\rr''),
\end{equation}
and 
\begin{equation}\label{or_vanhove}
G(\rr',\rr'',t) = \frac{1}{\rho_b} \left<   \delta(\rr''-\rr) e^{t  \mathcal{L}_{d}^\dagger } 
\delta(\rr'-\rr) \right>^{\rr}_{eq}
\end{equation}
is similar to a Van Hove function, but with a space-dependent diffusion constant 
\cite{vanhove1954correlations,hansen1990theory}.
In the case of a space-independent diffusion tensor, Eq. \eqref{or_vanhove} would become a Van Hove function
after integrating out one of the coordinates.

By working out what $\nabla_{\rr} \cdot \Gamma^{-1}(\rr)$ is want finds that there is no orientation along
the direction of the magnetic field.
Working out the derivative gives
\begin{align}
\nabla_{\rr} \cdot \Gamma^{-1}(\rr) =
    -\frac{1-\kappa(\rr)^2}{\left[1+\kappa(\rr)^2 \right]^2}
        \nabla_{\rr} \kappa(\rr) \cdot M \nonumber \\
    +\frac{2\kappa(\rr)\kappa'(\rr)}{\left[1+\kappa(\rr)^2\right]^2}
        \nabla_{\rr} \kappa(\rr) \cdot M^2.
\end{align}
Without loss of generality one can consider the case where the magnetic field is oriented along the $z$ direction.
In this case the matrices $M$ and $M^2$ are
\begin{align}
M = 
     \begin{bmatrix}
     0 & -1 & 0 \\
     1 & 0 & 0 \\
     0 & 0 & 0
    \end{bmatrix}
   \text{~and~}
   M^2=
       \begin{bmatrix}
    -1 & 0 & 0 \\
     0&-1&0 \\
     0&0&0
    \end{bmatrix},
\end{align}
so clearly there is no orientation along the direction of the magnetic field.

The operator $\mathcal{L}_{d}^\dagger$ acting on $\rr$ in Eq. \eqref{or_plin_r''} has the same effect as acting with
the adjoint on $\rr'$ (see, for example, Chap. 4.2 of Ref. \cite{risken1989fokker}); therefore
\begin{align}\label{or_p_avg_adj}
 \ppb^{(1)}(\rr') = - \ff& \int_0^\infty dt  \frac{e^{-2D_r t}}{3 \rho_b\gamma} \int d\rr'' 
 \left[ \nabla_{\rr''} \cdot \Gamma^{-1}(\rr'') \right] \nonumber \\
 & \left<  \delta(\rr''-\rr) e^{t \mathcal{L}_{d}(\rr') }
  \delta(\rr'-\rr) \right>^{\rr}_{eq}.
\end{align}
where
 \begin{equation}
 \mathcal{L}_{d}(\rr') =  T \nabla_{\rr'}  \cdot  \Gamma^{-1}(\rr') \cdot \nabla_{\rr'},
\end{equation}
and 
\begin{align}
 \frac{1}{\rho_b} \left< \delta(\rr''-\rr)\delta(\rr'-\rr) \right>^{\rr} = \delta(\rr''-\rr') \frac{\rho(\rr')}{\rho_b},
\end{align}
was used.
In the linear-response calculation of the orientation $\rho(\rr')$ can be replaced with $\rho_b$.

Equation \eqref{or_p_avg_adj} can be simplified to
\begin{align}
 \ppb^{(1)}(\rr') = &- \ff \int_0^\infty dt  \frac{1}{3\gamma}e^{-2D_r t}  e^{t \mathcal{L}_{d}(\rr') }   \nonumber \\
 \int d\rr'' 
 &   \delta(\rr''-\rr') \nabla_{\rr''} \cdot \Gamma^{-1}(\rr'') ,
\end{align}
and after the $\rr''$ integral this becomes
\begin{align}
 \ppb^{(1)}(\rr) = - \ff \int_0^\infty dt  \frac{1}{3\gamma}e^{-2D_r t}   
  e^{t \mathcal{L}_{d}(\rr) }  \nabla_{\rr} \cdot \Gamma^{-1}(\rr).
\end{align}
Multiplying both sides of this equation by $2D_r-\mathcal{L}_{d}(\rr)$ results in
\begin{align}
\left[ 2D_r-\mathcal{L}_{d}(\rr) \right] & \ppb^{(1)}(\rr) = \frac{\ff}{3\gamma} \int_0^\infty dt  \nonumber \\
& \frac{\partial}{\partial t} e^{-t \left[ 2D_r -\mathcal{L}_{d}(\rr) \right] }  
  \nabla_{\rr} \cdot \Gamma^{-1}(\rr),
\end{align}
which is equivalent to the following differential equation:
\begin{align}\label{or_plin_eq}
\left[ 2D_r-\mathcal{L}_{d}(\rr) \right]  \ppb^{(1)}(\rr) = -\frac{\ff}{3\gamma}  
  \nabla_{\rr} \cdot \Gamma^{-1}(\rr).
\end{align}
This equation is the main result of this appendix.
Solutions can be obtained numerically, or by further approximations.

For large $D_r$ and a magnetic field with small gradients, the contribution from the operator $\mathcal{L}_d$
to the left hand side of  Eq. \eqref{or_plin_eq} is negligible compared to the contribution coming from the 
part with $D_r$; therefore, one can approximate, in this case, the solution to this equation by
\begin{align}\label{or_plin_eq_1st_order}
 \ppb^{(1)}(\rr) \approx -\frac{1}{3} l  
  \nabla_{\rr} \cdot \Gamma^{-1}(\rr).
\end{align}
Note that this is the same
Eqs. \eqref{ge_px} and \eqref{ge_py} without the terms of cubic order in $\ff$.

Even though the density does not have a linear response to the swim speed,
the linear response calculations for the orientation can be used to calculate the density and fluxes \cite{sharma2017brownian}.
First, one integrates out the orientational degrees of freedom from Eq. \eqref{sml_FPE_Q_flux}:
\begin{align}
\rho(\rr,t) = - \nabla_{\rr} \cdot \jj(\rr,t),
\end{align}
with
\begin{align}\label{linres_j}
\jj(\rr,t) = - \frac{1}{\gamma} \Gamma^{-1} \cdot
\left[ T \nabla_{\rr} - \ff \ppb(\rr,t) \right] \rho(\rr,t),
\end{align}
where $\ppb(\rr,t) \equiv \frac{1}{\rho(\rr,t)} \int d \pp Q(\rr,\pp,t)$ is the polarization.

When the magnetic field is oriented in the $z$ direction and depends only on the $x$ coordinate, equating the steady-state flux in the $x$ direction to zero gives the following density:
\begin{align}\label{linres_rho}
\rho^{(2)}(x) \propto \exp\left[ \frac{\ff}{T} \int^x
dx' \pb^{(1)}_x(x') + \kappa(x') \pb^{(1)}_y(x') \right],
\end{align}
where the proportionality constant is determined by normalization, and the superscript $(2)$ indicates that it is correct up to second order in the swim force (there is no linear-order response in the density).
With this density the steady-state flux in the $y$ direction can be calculated from Eq. \eqref{linres_j}:
\begin{align}\label{linres_jy}
j^{(2)}_y(x) = \frac{\rho_b \ff}{\gamma} \pb^{(1)}_y(x),
\end{align}
where $\rho_b$ is the bulk density.

\section{Improved solution}\label{app:improved_solution}

\begin{figure}[b]
\includegraphics[width=1.\columnwidth]{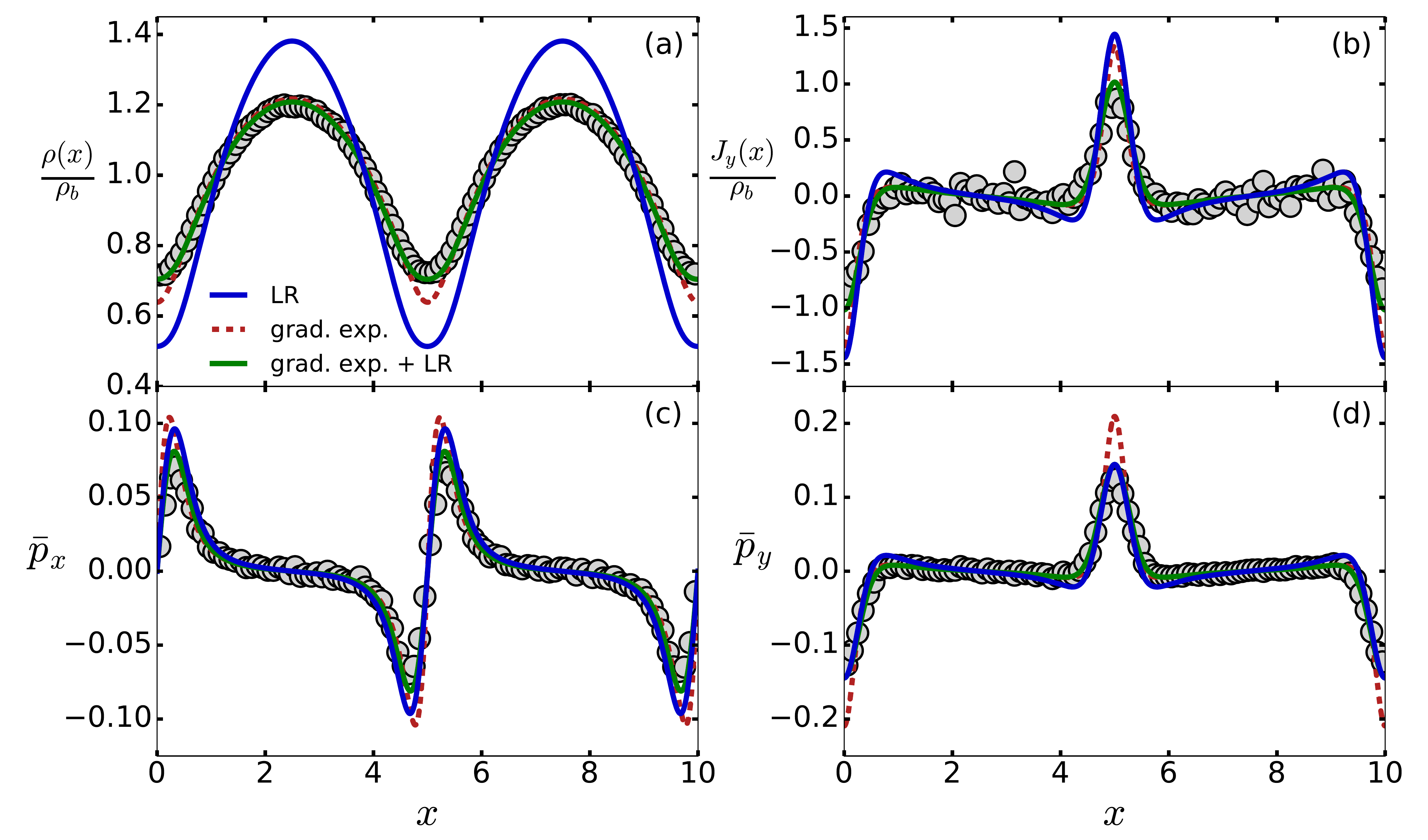}
\caption{Same as Fig. \ref{fig:bulk_results}, but with the linear-response solution (blue).}\label{fig:linear_response}
\end{figure}

Only using the linear order contributions in $f$ to the orientation gives poor agreement with simulations for the density; see Fig. \ref{fig:linear_response}.
The terms of cubic order in $f$ contribute to the diffusion in the system,
and by ignoring them there has to be a larger density gradient to compensate for the flux resulting from the orientation.
This problem can be overcome by combining the first order result from linear-response theory with the cubic terms from the gradient expansion.

Using the gradient expansion (App. \ref{app_gradient_expansion}) we found that the flux is
\begin{align}\label{imp_flux}
\jj(\rr,t) = - \Gamma^{-1} \cdot
	\left[ \ff \rho \ppb  \right]  
- \frac{T}{\gamma} \Gamma^{-1} \cdot \nabla \rho,
\end{align}
where $\ppb=\ppb(\rr,t)$ is the average polarization at position $\rr$, which is
\begin{align}
\ppb(\rr,t) &= -\frac{l}{3\rho} \nabla \cdot 
\left( \Gamma^{-1} \rho \right) \nonumber \\
&=
    -\frac{l}{3} \nabla \cdot \Gamma^{-1}
    - \frac{l}{3} \Gamma^{-1,T} \cdot
        \nabla \ln \rho.
\end{align}
The second term on the right hand side is higher order in the swim force because a gradient in the density is at least second order.
The first term on the right hand side of this equation is proportional to the swim force and is equal to the linear-response solution in the limit of small gradients.
However, Eq. \eqref{or_plin_eq} gives the linear-order response of the polarization without assuming that the gradients are small.
We therefore replace the linear order term with the numerical solution to the linear-response equation for the polarization [Eq. \eqref{or_plin_eq}] and rewrite the previous equation as
\begin{align}\label{imp_p_split}
\ppb(\rr,t) = \ppb^{(1)} -
 \frac{l}{3} \Gamma^{-1,T} \cdot
        \nabla \ln \rho.
\end{align}
When the magnetic field is oriented in the $z$ direction and varies in the $x$ direction, one can use this expression for the polarization together with the expression for the flux
[Eq. \eqref{imp_flux}] to obtain the steady-state density by equating the flux in the $x$ direction to zero:
\begin{align}\label{imp_rho}
\rho(x) = N \exp \left[ \frac{f}{T+\Delta T}
\int^x_0 dx' \pb_x^{(1)}(x') +
\kappa(x) \pb_y^{(1)}(x')
\right],
\end{align}
where $N$ is a normalization constant such that $\int_0^L dx \rho(x) = 1$ and $L$ the size of the periodic box.
Note that this expression is the same as the density calculated only using linear-response theory [Eq. \eqref{linres_rho}], but with $T$ replaced by $T+\Delta T$.
The flux in the $y$ direction can be calculated from Eq. \eqref{imp_flux} together with the result from the density as
\begin{align}
j_y(x) = \frac{\ff}{\gamma} \rho \pb_y^{(1)} +
\frac{f\delta}{\gamma}
\frac{\kappa}{1+\kappa^2} \left(
\pb_x^{(1)} - \kappa \pb_y^{(1)} \right).
\end{align}
And finally, the density can be used in Eq. \eqref{imp_p_split} to obtain the polarization in the $x$ and $y$ directions in terms of the linear-response polarization as 
\begin{align}
\pb_x(x)= \pb_x^{(1)} - 
\frac{\delta}{1+\kappa^2}
\left( \pb_x^{(1)} +\kappa \pb_y^{(1)}
\right),
\end{align}
and
\begin{align}
\pb_y(x)= \pb_y^{(1)} - 
\frac{\delta \kappa}{1+\kappa^2}
\left( \pb_x^{(1)} +\kappa \pb_y^{(1)}
\right).
\end{align}

\section{simulation details and numerics}\label{app:numerics}

The Brownian dynamics simulations were done by using the Euler algorithm to discretize the equations of motion in time [Eqs. \eqref{sde_r}-\eqref{sde_p}].
After each time increment of the orientation vector [Eq. \eqref{sde_p}], the vector is rescaled such that the length is always unity.
For all simulations a time step of $\Delta t = 0.5\ 10^{-6}$ was used.
The flux in the system was obtained by sampling the velocity of the ABP.
The linear-response equation for the orientation [Eq.  \eqref{or_plin_eq}] was integrated using the boundary value problem solver from the Scipy Python library~\cite{scipy}.

\bibliographystyle{apsrev4-1}


%

\end{document}